\documentclass[preprint,12pt]{elsarticle}

\usepackage{hyperref}

\makeatletter
\def\ps@pprintTitle{%
	\let\@oddhead\@empty
	\let\@evenhead\@empty
	\let\@oddfoot\@empty
	\let\@evenfoot\@oddfoot
}
\makeatother

\biboptions{authoryear}

\usepackage{natbib}
\usepackage{amsmath, amssymb, amsthm,commath,mathtools}
\usepackage{graphicx}
\usepackage{subcaption}
\usepackage{dsfont}
\usepackage{changepage}
\usepackage{float,relsize}
\usepackage{textcomp}
\usepackage{enumitem}
\usepackage{algorithm}
\usepackage{algpseudocode}

\newcommand{\bE}[1]{\mathds{E}\left \{ #1 \right\}}

\newcommand{\bb}[1]{\textbf{#1}}
\newcommand{\co}[2]{#1\;\lvert\rvert\; #2}

\newcommand*{\chaindash}{\,\rule[0.5ex]{1em}{0.55pt}\,}
\newcommand{\emp}[0]{\textnormal{emp}}
\newcommand{\ct}[0]{\textnormal{CT}}
\newcommand{\trh}[0]{\textnormal{th}}
\newcommand{\nor}[0]{\textnormal{norm}}

\newcommand{\cov}[0]{\textnormal{cov}}

\newcommand{\cg}[1]{{\color{gray} #1}}

\newtheorem{assumption}{\bf Assumption}
\newtheorem{theorem}{\bf Theorem}

\usepackage{tikz}

\begin{document}

\begin{frontmatter}

\title{Causality Graph of Vehicular Traffic Flow}

\author[]{Sina~Molavipour\corref{mycorrespondingauthor}}
\cortext[mycorrespondingauthor]{Corresponding author}
\ead{sinmo@kth.se}

\author{Germ\'{a}n~Bassi}

\author{Mladen~\v{C}i\v{c}i\'{c}}

\author{Mikael~Skoglund}

\author{Karl~Henrik~Johansson}

\address{School of Electrical Engineering and Computer Science, KTH Royal Institute of Technology, Stockholm, Sweden}

\begin{abstract}
In an intelligent transportation system, the effects and relations of traffic flow at different points in a network are valuable features which can be exploited for control system design and traffic forecasting. In this paper, we define the notion of causality based on the directed information, a well-established data-driven measure, to represent the effective connectivity among nodes of a vehicular traffic network. This notion indicates whether the traffic flow at any given point affects another point's flow in the future and, more importantly, reveals the extent of this effect. In contrast with conventional methods to express connections in a network, it is not limited to linear models and normality conditions.

In this work, directed information is used to determine the underlying graph structure of a network, denoted \textit{directed information graph}, which expresses the causal relations among nodes in the network.
We devise an algorithm to estimate the extent of the effects in each link and build the graph. The performance of the algorithm is then analyzed with synthetic data and real aggregated data of vehicular traffic.
\end{abstract}

\begin{keyword}
Vehicular traffic network\sep Causal graph\sep Directed information
\end{keyword}

\end{frontmatter}

\begingroup\def\thefootnote{}\footnotetext{This work was supported in part by the Knut and Alice Wallenberg Foundation and the Swedish Foundation for Strategic Research.}\endgroup


\section{Introduction}
Traffic control systems heavily rely on our understanding of driving patterns and models which describe particular traffic features. Adequate inference enables us to have a better prediction of traffic flow, and thus shape it accordingly; this, in turn, improves the efficiency of the whole network and the experience for road users. 
That is the goal of modern intelligent transport systems (ITS).
For instance, in a vehicle routing service, the knowledge about an imminent congestion and its predicted impact on other local links can be considered in choosing the optimal route.
Furthermore, this type of information is also valuable in urban traffic planning to characterize the required public road infrastructure~\cite{keimer2018information}.
Private enterprises also benefit from leveraging information from ITS. In~\cite{besselink2016cyber} a cyber-physical transportation system is studied for a freight network; the knowledge of the underlying effects and causation patterns among nodes of such a network improves the efficiency of the system, e.g., by reducing fuel consumption.

%
%

There are many examples of measures reflecting different aspects of traffic data which have been adopted as inputs in ITS. The origin-destination matrix, for instance, is estimated in many macroscopic applications to capture traffic behaviors and is demanded in long-term planning (\citet{ma2017estimation, TYMPAKIANAKI2015231, ALEXANDER2015240}).
The spatio-temporal correlation, or a similar metric denoted coefficient of determination (CoD), is another valuable data-driven feature which has gained attention in recent studies since advances in traffic monitoring have provided adequate data for development of models and forecasting mechanisms. This measure is obtained by observing time series of traffic flows at different points and computing the empirical correlation (\citet{VLAHOGIANNI20143, cheng2012spatio, CAI201621, MIN2011606, ermagun2017using}). 
In particular, in~\cite{salamanis2016managing, diamantopoulos2013investigating}, CoD is employed to create spatio-temporal models which are shown to be better, compared to non-parametric models such as $k$-nearest neighbors and support vector machine, in terms of prediction power.
\citet{salamanis2016managing} investigate graph-based techniques to optimize computations of CoD for a whole network, which are used to build its spatio-temporal model. 
In addition to prediction, spatio-temporal measures can reveal information about the causal effects and structures.
The value and location of peaks in the cross-covariance function, which is empirically computed from the data, can be employed to identify causal relationships (\citet{cheng2012spatio}).
A positive/negative peak indicates an excitation/inhibition, while the location determines the time shift that the effect needs to propagate. 
One limitation of these studies is that they are based on auto regressive models which work optimally only if the effects operate linearly.

The direction of the effect is paramount to describe the relationship between two nodes in a traffic network. Even in a one-way road, congestion can propagate backward and affect the future flow of preceding points. \citet{cheng2012spatio} discuss this backward effect in the traffic flow due to congestion in a scenario of merging roads. A more thorough study focused on congestion is done by~\citet{treiber2012validation} which is based on speed time series. The authors consider a wave model for back propagation of congestion and compute the parameters using the cross-covariance function and location of the peak.
In spatio-temporal correlation analysis, by fixing the time lag, the obtained cross-covariance matrix (expressed as a spatial weight matrix) is symmetric; thus, the network is described by a undirected graph. Nonetheless, the physical direction of the road can be enforced to create the appropriate adjacency matrix and obtain a directed graph (\citet{KAMARIANAKIS2005119}).
Furthermore, although the location of the peaks can indicate the direction of the effect for a specific time shift, it is not trivial to determine the extent of the effect in either direction. In this paper we try to answer the following question: \textit{how much does the traffic in one point causally affect the traffic in another point?}

The debate regarding how to quantify causal effects between two time series has been around for many years in statistics, and different notions have been suggested for causality. \citet{granger1969investigating} compares the variance of the noise between two possible models to identify the causal effects while \citet{PIERCE1977265} investigate the relationship between causality and cross-covariance. In both studies, the analysis is tied to enforce linear models and normality on the time series to perform optimally. Moreover, the extent of the causal influence is not clear to measure. Subsequently, Granger's notion is extended with the likelihood criterion (\citet{Rissanen1987}) and the discussion on inferring causality is taken by the community of information theory. The notions of directed information and transfer entropy are then introduced to measure causality, which is no longer restricted to linear models while still consistent with Granger's notion and cross-covariance metrics (see~\cite{quinn2011estimating} for a review). 

In order to graphically describe the structure of causal effects in a network, \citet{quinn2011estimating, amblard2011directed} introduce and justify directed information graphs (DIG), which is the motivation of our study. This model is also adopted and used to analyze a wide variety of different applications such as economy~\citet{jiao2013universal}, neuroscience~\citet{cai2017inferring}, and social networks~\cite{quinn2015directed}. \citet{jiao2013universal} and \citet{quinn2011estimating} address the problem of estimating the directed information by proposing a context-tree based and plug-in estimator, respectively. The performance of detecting the DIG is investigated as a hypothesis test problem in~\citet{kontoyiannis2016estimating, Mol2017TestforDIG}.

In this paper, we present and motivate the use of an algorithm that identifies the underlying causal structures of vehicular networks. The algorithm is oblivious to the true physical structure of the network and it works by estimating the DIG from aggregated data of vehicular flow.
The main purpose of this work is to introduce a new measure for causal dependencies in traffic networks which can be used alongside more traditional model-based approaches.
The rest of the work is organized as follows.
In Section~\ref{sec:methods}, we first review the preliminaries and definition of directed information graph and we proceed by outlining the steps of our method. 
We present two different estimators for the directed information and we elaborate on some relevant aspects to reduce their complexity; the memory of the estimator, which is related on how much past information should be analyzed, and the quantization of data prior to the estimation are two important considerations.
Next, we validate our approach by applying it to different scenarios in Section~\ref{sec:results}. The evaluation is performed on synthetic data, modeled by either a Poisson distribution or the cell transmission model (CTM), and on real data aggregated from the California Department of Transportation. Finally the paper is concluded in Section~\ref{sec:concl}.

\section{Methods}
\label{sec:methods}

\subsection{Notation}
For two integers $i$ and $j$, ${i:j}$ indicates the sequence $i,i+1,\dots,j$. For two sets $S$ and $Q$, $S\setminus Q$ denotes the difference of set $Q$ and $S$. Moreover, $\abs{S}$ indicates the size of the set. 

For a matrix $A$ with elements $A_{ij}$, $\abs{A}$ denotes the matrix with elements $|A_{ij}|$. The normalized matrix is then defined as $A_\nor\triangleq{\abs{A}}/{\max\{\abs{A}\}}$. 

For a random process \bb{X}, the randomly generated time series with $n$ elements is expressed by $X_1^n=X_1,\dots,X_n$, or simply $X^n$, and $x_i\in\mathcal{X}$ denotes the realization of the $i$-th sample.
If the process is indexed with $(m)$, i.e., $\bb{X}_{(m)}$, we use $X^n_{(m)}$ to show the time series. A random process corresponds to a node in the graph representation of a network.

For two random variables $X$ and $Y$, the mutual information between them is expressed as $I(X;Y)$, and $H(X)$ stands for the entropy of $X$.

\subsection{Vehicular traffic flow}

Vehicular flow conventionally denotes the number of cars passing by a specific point on the road per unit of time. Among parameters to monitor traffic, the flow of vehicles plays an important role in controlling the behavior of the system. Many aspects of the traffic at a certain point, such as congestion or patterns of rush hour, can be captured directly from the data of one sensor. However, as the transportation infrastructure is physically connected, the traffic flow at one point can effectively influence another point in a local area. Although the effect of a direct physical link could be expressed with models, it seems non-trivial to capture indirect links in spite of their possibly dominant influence. Furthermore, models would get complicated as the scale changes from microscopic to mesoscopic and macroscopic.

By analyzing time series data from different sensors, we can infer statistical influences among sensors on a network. A conventional undirected graph which reveals the connection between two points is the correlation graph, in which an undirected link between two nodes exists if they are correlated. In this work, we are interested in testing if the signal at one sensor is causally controlling the signal at another sensor. For instance, if the highway is in free flow, the changes in flow are sensed with a delay at a downstream location, while in a case of congestion the effect propagates backward.
Directed information is a well-established information-theoretic measure to test the existence of such causal links. Based on this notion, the concept of directed information graph (DIG) is proposed; in the following, we review its characteristics.

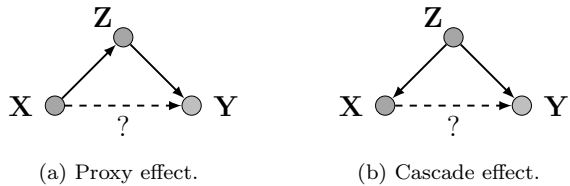
\begin{figure}[t]
\centering
\begin{subfigure}[t]{0.25\textwidth}
	\resizebox{\columnwidth}{!}{
	\begin{tikzpicture}[]
		\node[shape=circle,draw=black,fill=gray!70,inner sep=0pt,minimum size=8pt] (X) at (0,0) {};
		\node (XT) at (-0.5,0) {$\bb{X}$};
		\node[shape=circle,draw=black,fill=gray!70,inner sep=0pt,minimum size=8pt] (Z) at (1,1) {};
		\node (XT) at (0.7,1.3) {$\bb{Z}$};
		\node[shape=circle,draw=black,fill=gray!50,inner sep=0pt,minimum size=8pt] (Y) at (2,0) {};
		\node (XT) at (2.5,0) {$\bb{Y}$};
		\draw [-latex,thick] (X) -- (Z);
		\draw [-latex,thick] (Z) -- (Y);
		\draw [-latex,thick,dashed] (X) --node [midway,below]{?} (Y);
	\end{tikzpicture}}
	\caption{Proxy effect.}
\end{subfigure}
\qquad\begin{subfigure}[t]{0.25\textwidth}
	\resizebox{\columnwidth}{!}{
	\begin{tikzpicture}[]
		\node[shape=circle,draw=black,fill=gray!70,inner sep=0pt,minimum size=8pt] (X) at (0,0) {};
		\node (XT) at (-0.5,0) {$\bb{X}$};
		\node[shape=circle,draw=black,fill=gray!70,inner sep=0pt,minimum size=8pt] (Z) at (1,1) {};
		\node (XT) at (0.7,1.3) {$\bb{Z}$};
		\node[shape=circle,draw=black,fill=gray!50,inner sep=0pt,minimum size=8pt] (Y) at (2,0) {};
		\node (XT) at (2.5,0) {$\bb{Y}$};
		\draw [-latex,thick] (Z) -- (X);
		\draw [-latex,thick] (Z) -- (Y);
		\draw [-latex,thick,dashed] (X) --node [midway,below]{?} (Y);
	\end{tikzpicture}}
	\caption{Cascade effect.}
\end{subfigure}

\caption{Effects of latent node $\bb{Z}$ on causal link between $\bb{X}$ and $\bb{Y}$.}
\label{fig:effects}
\end{figure}

\subsection{Directed information graph}
\label{sec:DIG}
Consider three random processes $\bb{X},\bb{Y}$, and $\bb{Z}$; then the \emph{directed information rate} from $\bb{X}$ to $\bb{Y}$ \emph{causally conditioned} on $\bb{Z}$ is defined as: 
\begin{align}
I(\co{\bb{X}\to \bb{Y}}{\bb{Z}})&\triangleq \lim\limits_{n\to\infty}\frac{1}{n}\sum_{i=1}^{n}I(Y_{i}\,;\, X_{1}^{i}|Y_{1}^{i-1},Z_{1}^{i}). \label{eq:DI_def}
\end{align}
The definition in~\eqref{eq:DI_def} is based on non-strictly causal dependency. In this sense, instantaneous effects between two nodes appear in both causal directions; in other words, both $I(\co{\bb{X}\to \bb{Y}}{\bb{Z}})$ and $I(\co{\bb{Y}\to \bb{X}}{\bb{Z}})$ share the common terms $I(Y_{i}\,;\, X_{i}|X_{1}^{i-1},Y_{1}^{i-1},Z_{1}^{i})$, $\forall i\in\{1:n\}$, by the chain rule of the mutual information. 
Without considering the effect of $\bb{Z}$, a causal link exists from $\bb{X}$ to $\bb{Y}$ if and only if $I(\bb{X}\to \bb{Y})>0$ (given that the mutual information is always non-negative). However, the signal from \bb{Y} may become causally independent of $\bb{X}$ if the signal from $\bb{Z}$ is known.

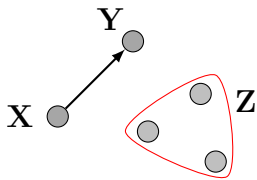
\begin{figure}
\centering
\begin{tikzpicture}
\node[shape=circle,draw=black,fill=gray!70,inner sep=0pt,minimum size=8pt] (X) at (0,0) {};
\node (XT) at (-0.5,0) {$\bb{X}$};
\node[shape=circle,draw=black,fill=gray!70,inner sep=0pt,minimum size=8pt] (Y) at (1,1) {};
\node (XT) at (0.7,1.3) {$\bb{Y}$};
\node[shape=circle,draw=black,fill=gray!50,inner sep=0pt,minimum size=8pt] (Z1) at (1.2,-0.2) {};
\node[shape=circle,draw=black,fill=gray!50,inner sep=0pt,minimum size=8pt] (Z2) at (1.9,0.3) {};
\node[shape=circle,draw=black,fill=gray!50,inner sep=0pt,minimum size=8pt] (Z2) at (2.1,-.6) {};
\node (XT) at (2.5,0.2) {$\bb{Z}$};
\draw [red] plot [smooth cycle,dashed] coordinates {(0.9,-0.2)  (2.1,0.55)  (2.25,-.8) };
\draw [-latex,thick] (X) -- (Y);
\end{tikzpicture}
\caption{Causally conditioned directed information between $\bb{X}$ and $\bb{Y}$ requires the history of the hyper-node $\bb{Z}$.}
\label{fig:hyperNode}
\end{figure}

In fact there are two possible cases where the knowledge of $\bb{Z}$ can change the causal relation from $\bb{X}$ to $\bb{Y}$: the \textit{proxy effect} (Figure~\ref{fig:effects}.a) and the \textit{cascade effect} (Figure~\ref{fig:effects}.b). In the first scenario, information does flow from $\bb{X}$ to $\bb{Y}$ but not directly; thus, by causally conditioning on $\bb{Z}$, the directed information from $\bb{X}$ to $\bb{Y}$ becomes zero.
However, without conditioning, $I(\bb{X}\to \bb{Y})$ is positive and we would wrongly assume that there is a direct causation from $\bb{X}$ to $\bb{Y}$.
In the second scenario, the node $\bb{Z}$ affects both two other nodes,which causes a statistical correlation between them. If the effect on $\bb{X}$ appears before the one in $\bb{Y}$, we will detect a positive $I(\bb{X}\to \bb{Y})$ and we would again incorrectly assume that a causal link exists between them.
These two examples motivate the notion of directed information graph in which there exists a causal link from $\bb{X}$ to $\bb{Y}$ if and only if $I(\co{\bb{X}\to \bb{Y}}{\bb{Z}})>0$. 
Extending this definition to larger networks requires the assumption that $\bb{Z}$ is a hyper-node (collection of several nodes) which represents the whole network excluding $\bb{X}$ and $\bb{Y}$ (Figure~\ref{fig:hyperNode}). 
To detect edges in a DIG, the causally conditioned directed information is estimated for each pair of nodes and a threshold test on the value indicates the existence of directed links. In the following section, we review methods for estimating the directed information.

\subsection{Estimation}

Various techniques have been suggested to estimate information-theoretic measures such as entropy, and mutual and directed information. In this paper, we focus on two methods in which the probability distribution is estimated and subsequently used to compute the measure: the \emph{empirical} and \emph{context tree} (CT) estimators.
In general, the joint distribution $P(X^i,Y^i,Z^i)$ for any $i\in\{1: n\}$ is needed in order to estimate~\eqref{eq:DI_def}, where $n\to\infty$. 
The following assumptions ensure consistency of the estimation while they make the computations feasible.

\begin{assumption}
\label{asm:Markov}
For a graph with three nodes $(\bb{X},\bb{Y},\bb{Z})$ the following should hold true:
\begin{enumerate}[label=(\alph*)]
\item $\bb{X},\bb{Y}$, and $\bb{Z}$ are jointly stationary irreducible Markov of order $k$.
\item All transition probabilities $Q(X_{k+1},Y_{k+1},Z_{k+1}|X^{k},Y^{k},Z^{k})$ are positive.
\item For any pairs of nodes such as $\bb{X}$ and $\bb{Y}$, the Markov chain
\begin{align}
\bar X_{i} \chaindash (\bar X_{i-k}^{i-1}\bar Y_{i-k}^i) \chaindash (\bar X_{1}^{i-k-1}\bar Y_{1}^{i-k-1}) \nonumber
\end{align} must hold for $k<i\leq n$.
\end{enumerate}
\end{assumption}

If Assumption~\ref{asm:Markov} holds true, the directed information in~\eqref{eq:DI_def} can be simplified as:
\begin{align}
I(\co{\bb{X}\to \bb{Y}}{\bb{Z}})=I({Y_{k+1};X^{k+1}}|{Y^k,Z^{k+1}}), \label{eq:DI_Markov}
\end{align}
and estimating $P(X^{k+1},Y^{k+1},Z^{k+1})$ is sufficient.
Note that if only Assumption~\ref{asm:Markov}.a and \ref{asm:Markov}.b hold, the quantity in~\eqref{eq:DI_Markov} is in fact an upper bound for the directed information in~\eqref{eq:DI_def}:

\begin{align}
&I(\co{\bb{X}\to \bb{Y}}{\bb{Z}})=\lim\limits_{n\to\infty}\frac{1}{n}\sum_{i=k+1}^{n}I(Y_{i}\,;\, X_{1}^{i}|Y_{1}^{i-1},Z_{1}^{i}) \nonumber\\
&\quad= \lim\limits_{n\to\infty}\frac{1}{n}\sum_{i=k+1}^{n} \left[H(Y_{i}|Y_{1}^{i-1},Z_{1}^{i}) - H(Y_{i} |X_{1}^{i}, Y_{1}^{i-1},Z_{1}^{i})\right] \nonumber\\
&\quad=  I({Y_{k+1};X^{k+1}}|{Y^k,Z^{k+1}})- \lim\limits_{n\to\infty}\frac{1}{n}\sum_{i=k+1}^{n}I(Y_{i};Y_{1}^{i-k-1},Z_{1}^{i-k-1}|Y_{i-k}^{i-1},Z_{i-k}^{i})\nonumber\\
&\quad\leq I({Y_{k+1};X^{k+1}}|{Y^k,Z^{k+1}}).
\end{align}

\subsubsection{Plug-in empirical estimator}

In this method, the distribution $P(X^{k+1},Y^{k+1},Z^{k+1})$ is estimated empirically, by counting the patterns in the observation.
Then by plugging-in the empirical distribution into~\eqref{eq:DI_Markov}, we obtain an estimate of the directed information, denoted $\hat{I}_\emp(\co{\bb{X}\to \bb{Y}}{\bb{Z}})$.

For a pair of nodes, the plug-in estimator $\hat{I}_\emp(\bb{X}\to \bb{Y})$ is shown to be consistent with probability one (almost surely) when the source is stationary ergodic and both $(\bb{X},\bb{Y})$ and $\bb{Y}$ are Markov sources~\cite[Thm.~1]{quinn2011estimating}. The extension to networks with more than two nodes implies that the plug-in estimator $\hat{I}_\emp(\co{\bb{X}\to \bb{Y}}{\bb{Z}})$ is consistent with probability one if $(\bb{X},\bb{Y},\bb{Z})$ and $(\bb{Y},\bb{Z})$ are ergodic stationary Markov sources of the same order (Assumption~\ref{asm:Markov}).

To test a graph structure, it is required to perform a threshold test on all links, i.e., the estimated conditional directed information should be above a predefined threshold $I_\trh$. We say the graph is detected correctly if the adjacency matrix of the estimated graph is equal to the one of the true underlying graph. The performance of such a test is addressed in the following theorem. It indicates that using the empirical estimator, the type I and type II errors of testing DIG are asymptotically zero, given a proper choice of $I_\trh$.

\begin{theorem}[\citet{Mol2017TestforDIG}]
Consider a network of $M$ sensors. For a directed information graph with adjacency matrix $V$ of size ${M\times M}$, if Assumption~\ref{asm:Markov} holds, the performance of the test for the hypothesis $V^*$ is bounded as:
\begin{align}
P_F &\leq 1-P_G\left(\frac{R}{2},I_{th}\right), \nonumber\\
P_D &\geq \max\!\left\{1-W_0\!\left[1-P_G\left(\frac{R}{2},I_{th}\right)\right],0\right\}, 
\end{align}
using the plug-in estimation of $n$ samples with $n\to\infty$. The function $P_G$ is the \emph{regularized gamma function}, and $W_0=M(M-1)-W_1$ with $W_1$ denoting the number of directed edges in the hypothesis graph, and $R=\abs{\mathcal{X}}^{M\,k}( \abs{\mathcal{X}}^{M}-1)$. Finally, $I_{th}$ is the threshold value used to decide the existence of an edge, and its order is $\mathcal{O}(1)$.
\end{theorem}

\subsubsection{Context tree estimator}

Although the empirical estimator is consistent in evaluating the distribution, as the dimension of the model increases, it requires more samples to achieve good estimates. Consequently, in practice, it is crucial to reduce the computational complexity of the estimation. In addition, by the Markov assumption (Assumption~\ref{asm:Markov}.a), we consider all patterns of sequence with a depth $k$, while in practice some patterns may rarely appear compared to others, and computing all possible patterns would not be efficient. To address this, the context tree algorithm was proposed by~\citet{willems1995context,willems1998context} for a class of stationary ergodic finite-alphabet sources, and shown to be linear in the number of samples ($n$).
In~\citet{jiao2013universal}, the estimation of the directed information based on CT is investigated and the consistency is assured with probability one, as long as the source is irreducible aperiodic Markov (Assumption~\ref{asm:Markov}).

The estimated directed information can be computed in several ways (\citet{jiao2013universal}); each estimator has a different convergence speed and behavior with respect to $n$. In our method, we have chosen our CT estimator to be 
\begin{align}
\hat I_\ct(\co{\bb{X}\to \bb{Y}}{\bb{Z}})\triangleq \frac{1}{n} \sum_{i=1}^{n} D\!\left(\co{\hat{P}_\ct(y_i|X^i,Y^{i-1},Z^i)}{\hat{P}_{CT}(y_i|Y^{i-1},Z^i)}\right),
\label{eq:DI_est}
\end{align}
where $D(\co{\cdot}{\cdot})$ is the relative entropy where we average over different choices of $y_i$,
and $\hat{P}_\ct(\cdot)$ is the estimated distribution based on the context tree method. Note that the estimator is non-negative due to the non-negativity of the relative entropy.

\subsection{Estimation of memory depth}
\label{Subsec:Est_depth}

For both estimators, the maximum memory depth $d$ for which the causal effects are taken into consideration--through $\hat{P}(X^{d+1},Y^{d+1},Z^{d+1})$--needs to be determined. If the source is Markov, one can fix any memory greater than the order of the Markov process. However, in real data where the process is unknown, determining the memory depth is not trivial. In simple traffic scenarios where the physical location of the sensors and average speeds of the vehicles are available, an approximate depth can be calculated. Nevertheless, our approach to determine the memory depth is data-driven, depends only on the traffic flows, and it is based on statistics of signals from each sensor.

Motivated by spatio-temporal analysis, for any pair of nodes $\bb{X}$ and $\bb{Y}$, we evaluate the cross-covariance between traffic flows and the location of the peak determines how much delay is required such that the signals become statically correlated:
\begin{align}
\cov_{X,Y}(\tau)&=\frac{1}{n-\tau}\sum_{i=1}^{n-\tau}(X_{i+\tau}-\mu_X)(Y_i-\mu_Y)\qquad \tau\in\{0,\dots,n-1\}\\
d_{XY}&=\underset{\tau}{\operatorname{arg\,max}} \{\cov_{X,Y}(\tau)\},
\end{align}
where $\mu_X$ and $\mu_Y$ are the sample means.
Then we choose the depth $d$ for estimation to be the maximum among all computed delays.

Choosing an insufficient depth will likely cause an erroneous estimation with either of the estimators in this paper. In the sequel, we only discuss the effects on the plug-in empirical estimator due to its simplicity, and later comment on the CT estimator.

Consider a network with three nodes $\{\bb{X},\bb{Y},\bb{Z}\}$ which fulfills Assumption~\ref{asm:Markov}, and suppose we estimate the causal link from $\bb{X}$ to $\bb{Y}$.
Let us indicate the empirical estimator with the assumption of a $d$-th order Markov process as
\begin{equation}
\hat{I}_\emp^{(d)}(\co{\bb{X}\to \bb{Y}}{\bb{Z}})\triangleq \hat I_\emp({Y_{d+1};X_1^{d+1}}|{Y_1^d,Z_1^{d+1}}).
\end{equation}
Note that in the limit of large number of samples, this estimation is consistent.
If the estimated depth were $\hat d=k-1$, i.e., one order smaller than the true value, then the empirical plug-in estimator is given by
\begin{align}
\hat I^{(k-1)}_\emp(\co{\bb{X}\to \bb{Y}}{\bb{Z}})=\hat I_\emp({Y_{k+1};X_2^{k+1}}|{Y_2^k,Z_2^{k+1}}).
\end{align} 
Therefore, the (asymptotical) difference between $\hat I^{(k)}_\emp$ and $\hat I^{(k-1)}_\emp$ is:
\begin{align}
& \hat I_\emp({Y_{k+1};X_1^{k+1}}|{Y_1^k,Z_1^{k+1}})-\hat I_\emp({Y_{k+1};X_2^{k+1}}|{Y_2^k,Z_2^{k+1}})\nonumber\\
&\qquad= H(Y_{k+1}|Y_1^k, Z_1^{k+1})-H(Y_{k+1}|X_1^{k+1}, Y_1^k, Z_1^{k+1})\nonumber\\
&\qquad\quad  -H(Y_{k+1}|Y_2^k, Z_2^{k+1})+H(Y_{k+1}|X_2^{k+1}, Y_2^k, Z_2^{k+1})\label{eq:I_diff}\\
&\qquad= I(Y_{k+1};X_1,Y_1,Z_1|X_2^{k+1},Y_2^k,Z_2^{k+1})-I(Y_{k+1};Y_1,Z_1|Y_2^k,Z_2^{k+1}).\label{DI_d_diff}
\end{align}
In the case of the CT estimator, estimating $P(X_2^{k+1},Y_2^{k+1},Z_2^{k+1})$, i.e., the marginal of the joint probability corresponding to the $k$-th order Markov, is consistent only if we use the algorithm proposed in~\cite{willems1998context}. The proof of consistency for $\hat I_\ct^{(k-1)}$ requires a deeper analysis which is not the focus of this paper.

To understand how the trade-off in \eqref{DI_d_diff} behaves, consider the \emph{full network} $\{\bb{X},\bb{Y},\bb{Z}\}$ and the \emph{sub-network} $\{\bb{Y},\bb{Z}\}$ in the following extreme cases. If the \emph{sub-network} is Markov of order $k-1$ (while the whole network is of order $k$) the difference in \eqref{DI_d_diff} is positive, so the estimation of the directed information is always below the true value. On the other hand, \eqref{DI_d_diff} becomes strictly negative when the \emph{full network} is Markov of order $k-1$ and the \emph{sub-network} is not. Trivially, if the Markov property with order $k-1$ holds for both of them, the difference is zero.

Erroneous estimations are also possible in less extreme scenarios as we see next.
Consider three sensors $\bb{X}$, $\bb{Y}$, and $\bb{Z}$ in a network such that:
\begin{equation}\label{eq:lin_Pois_model}
\left\{
 \begin{array}{cl}
  X_i &= \,a_1\,Z_{i-1}+N_i,\\
  Y_i &= \,a_2\,X_{i-1}+a_3\,Z_{i}+N'_i,\\
  Z_i &= \,a_4\,Z_{i-2}+N''_i,
 \end{array}
\right.
\end{equation}
where $N$, $N'$, and $N''$ are independent random Poisson noises\footnote{%
The Poisson model is a conventional way to express links of traffic as queues (\citet{vandaele2000queueing}), and we have also used it to validate our method in Section~\ref{Sec:Poisson}.}; the coefficients in~\eqref{eq:lin_Pois_model} can be chosen to ensure stability of the system.
It is easy to see that the whole network and the sub-network $\{\bb{Y},\bb{Z}\}$ are both second order Markov processes, and thus our previous analysis regarding~\eqref{DI_d_diff} does not hold.
For this model, a consistent estimator of $I(\co{\bb{X}\to \bb{Y}}{\bb{Z}})$ should consider a depth $d=2$.
In the following, we show that with $n\to\infty$, estimators with smaller depths produce varying outputs, in particular:
\begin{equation}
\hat I_\emp^{(0)}(\co{\bb{X}\to \bb{Y}}{\bb{Z}}) \leq \hat I_\emp^{(2)}(\co{\bb{X}\to \bb{Y}}{\bb{Z}}) \leq \hat I_\emp^{(1)}(\co{\bb{X}\to \bb{Y}}{\bb{Z}}).
\label{eq:depth}
\end{equation}
In other words, by choosing the depth to be $d=1$ we get an upper bound for the correct estimator, while choosing $d=0$ yields a lower bound.

The right inequality in~\eqref{eq:depth} can be proved as follows
\begin{align}
&\hat I_\emp^{(2)}(\co{\bb{X}\to \bb{Y}}{\bb{Z}}) - \hat I_\emp^{(1)}(\co{\bb{X}\to \bb{Y}}{\bb{Z}})\nonumber\\
 &\hspace{.3cm}= H(Y_i|Y_{i-2}^{i-1},Z_{i-2}^i) -H(Y_i|X_{i-2}^i,Y_{i-2}^{i-1},Z_{i-2}^i) -H(Y_i|Y_{i-1},Z_{i-1}^i) \nonumber\\
 &\hspace{.7cm} +H(Y_i|X_{i-1}^i,Y_{i-1},Z_{i-1}^i)\nonumber\\
 &\hspace{.3cm}= H(a_2 N_{i-1} +N'_{i}) -H(N'_{i}) -H(a_1a_2Z_{i-2}+a_2N_{i-1}+N'_{i}| Y_{i-1}, Z_{i-1}^i) \nonumber\\
 &\hspace{.7cm} +H(N'_{i}) \nonumber\\
 &\hspace{.3cm}\leq H(a_2 N_{i-1} +N'_{i}) -H(a_2N_{i-1}+N'_{i}| a_2N_{i-2}+N'_{i-1}, Z_{i-2}^{i-1}, N''_{i}) \nonumber\\
 &\hspace{.3cm}=0,
\end{align}
where the inequality holds since conditioning on $N''_{i}$ reduces the entropy, and the final equality is due to $a_2 N_{i-1} +N'_{i}$ being independent of the quantities in the conditioning.
On the other hand, the left inequality in~\eqref{eq:depth} is trivial if we note that
\begin{align}
&\hat I^{(0)}(\co{\bb{X}\to \bb{Y}}{\bb{Z}})\nonumber\\
 &\hspace{.3cm}= I(X_i ;Y_i|Z_i) \nonumber\\
 &\hspace{.3cm}= I(a_1Z_{i-1}+N_{i} ;\, a_1a_2Z_{i-2}+a_2N_{i-1}+N'_{i}| a_4Z_{i-2}+N''_{i}) \nonumber\\
 &\hspace{.3cm}= 0.
\end{align}

The preceding analysis shows that, in this particular example, the choice of estimator depth may substantially change the final estimated graph.
Assume first that the coefficient $a_2$ is very small, which induces a small value of $I(\co{\bb{X}\to \bb{Y}}{\bb{Z}})$, and further assume that we are only interested in finding strong connections between nodes.
In this case, the choice $d=1$ may increase the estimated value of the directed information above the desired threshold, and thus we determine a causal connection $\bb{X}\to \bb{Y}$ when it was not the case.
On the contrary, if we are interested in detecting all causal connections and we choose $d=0$, we will fail to detect this relationship.

\subsection{Quantization}

The estimation cost is affected by the dimension of the data, in the sense of number of nodes and range of values that the vehicular flow can take. To picture the effect of range, note that the computational complexity of the empirical estimator is polynomial in the alphabet size of the flow values. 
To resolve this issue, the data of each sensor can be initially quantized to $r$ levels.
In consequence, we no longer estimate for the true traffic data, but a quantized value of that. Nevertheless, it makes sense as small variations in the number of cars have low impact on the overall flow of a link.
There is a trade-off in choosing $r$ between complexity and accuracy of the estimation which depends on the data and computational resources. Hereafter and with a slight abuse of terminology, we consider all traffic flows to be quantized.

To estimate the joint distribution of few random variables, we combine them into one with a larger alphabet. For example, $X$ and $Y \in \mathcal{X} \triangleq \{ 0,1,\dots,\abs{\mathcal{X}}-1 \}$ can be combined in a new random variable $W\triangleq\abs{\mathcal{X}}Y+X$ with alphabet size $\abs{\mathcal{X}}^2$.

\begin{algorithm}[t]
\caption{Detecting Directed Information Graph}
\label{alg:DIG}
\begin{algorithmic}[1]
 \Procedure{Estimate\_DIG}{$X_{(1)},X_{(2)},\dots,X_{(M)}$}
 	\State {$r \gets$ Fix number of quantization levels}
 	\State {$\alpha \gets$ Fix threshold for test}
    \For {every pair of $(m,l)$} 
     \State {$\cov_{m,l}$ $\gets$ cross\_covariance($X_{(m)},X_{(l)}$)} 
     \State {$d_{m,l}$ $\gets$ $|$position of peak of $\cov_{m,l}|$} 
    \EndFor
    \State {$d \gets$ $\max\limits_{m,l}$\{$d_{m,l}$\}}
    \For {every $m\in [1:M]$} 
     \State {$X_{(m)} \gets$ Quantize($X_{(m)},r$)}  
    \EndFor 
    \For {every pair of $(m,l)$} 
	 \State Compute $\hat P(X_{(m)}^{d+1},X_{(l)}^{d+1}|X_{\{1:M\}\setminus\{m,l\}}^{d+1})$ 
	 \State $I(m,l) \gets \hat I(\co{\bb{X}_{(m)}\to \bb{X}_{(l)}}{\bb{X}_{\{1:M\}\setminus\{m,l\}}})$ 	 
	 \State $H(m,l) \gets \hat H(\co{\bb{X}_{(l)}}{\bb{X}_{\{1:M\}\setminus\{m,l\}}})$ 	 
	 \State $G_{m,l} \gets \hat I(m,l)/H(m,l)$ 	 
    \EndFor
    \State {$G_\nor$ $\gets$ normalize $G$}
    \State {$DIG \gets (G_\nor\geq \alpha)$}
    \State\Return {G, DIG} 
  \EndProcedure
\end{algorithmic}
\end{algorithm}

\subsection{Test for the directed information graph}

The DIG can be created by applying a threshold test on the estimated values of directed information. It is however difficult to determine a proper value for the threshold given that we are a priori unaware of the range of values of the directed information, which may also vary depending on the quantization levels. Note that for any $n$, by the definition in~\eqref{eq:DI_est}, $\hat I_{CT}(\co{\bb{X}\to\bb{Y}}{\bb{Z}})\leq \hat H_{CT}(\co{\bb{Y}}{\bb{Z}})$ holds true where:
\begin{align}
\hat{H}_{CT}(\co{\bb{Y}}{\bb{Z}})&\triangleq -\frac{1}{n} \sum_{i=1}^{n} \sum_{y_i}\hat{P}_{CT}(y_i|X^i,Y^{i-1},Z^i)\log\left(\hat{P}_{CT}(y_i|Y^{i-1},Z^i)\right),\label{eq:H_ct}
\end{align}
and $\hat I_{\emp}(\co{\bb{X}\to\bb{Y}}{\bb{Z}})\leq \hat H_{\emp}(\co{\bb{Y}}{\bb{Z}})$ by the following definition:
\begin{align}
\hat{H}_{\emp}(\co{\bb{Y}}{\bb{Z}})&\triangleq -  \sum_{y^{k+1},z^{k+1}}\hat{P}_\emp(y^{k+1},z^{k+1})\log\left(\hat{P}_\emp(y_{k+1}|y^{k},z^{k+1})\right).
\label{eq:H_est}
\end{align}
As a result, we are able to normalize the causally conditioned directed information from $\bb{X}$ to $\bb{Y}$ for either of the estimators as:
\begin{equation}
\hat I^*_{X \to Y}\triangleq \frac{\hat I(\co{\bb{X}\to\bb{Y}}{\bb{Z}})}{\hat H(\co{\bb{Y}}{\bb{Z}})}, 
\label{eq:I_norm}
\end{equation}
where $\hat I$ and $\hat H$ correspond to the same estimator.

We can construct an adjacency matrix $G$ of the network using~\eqref{eq:I_norm}, where the off-diagonal entries are given by $\hat I^*_{X \to Y}$.
By the definition of the DIG, detecting links is performed by a threshold test on the entries of $G$. We may adjust the threshold to detect significant causal links in the graph or just the strongest ones. The detection steps are presented in Algorithm~\ref{alg:DIG}.


\section{Results and validations}
\label{sec:results}

In this section we evaluate our method in different scenarios both with simulated and real-world data.
The graph representation of the network (DIG) is obtained by computing the matrix $G$ through the values of the normalized directed information $\hat{I}^*$ as stated in Algorithm~\ref{alg:DIG}.
We further normalize the entire matrix $G$ (denoted as $G_\nor$), and we choose a threshold $0<\alpha\leq 1$ as the level of significance in order to distinguish the most important links. 
$\alpha$ is chosen such that we capture the links that are expected to exist (i.e., the links inferred from generative model in simulated scenarios, and the ones which follow the physical direction of the road in real data scenarios). Trivially by decreasing and increasing the threshold we can have type I and II errors in detecting a link. $\alpha$ can also be determined automatically by evaluating entries of $G_\nor$ which is not the focus of this paper.

In the following, we first simulate two common situations in road networks: the scenario where sensors are located along a single road and the scenario in which two roads merge into one. 
The simulations are performed in MATLAB and the road traffic is modeled either as a queue with Poisson input (Section~\ref{Sec:Poisson}) or with Cell Transmission Model (CTM) (Section~\ref{Sec:CTM}).
Then we analyze real data collected by installed sensors in highways and show how directed information can explain the direction of influence (Section~\ref{Sec:real}).

\subsection{Synthetic data (Poisson model)}
\label{Sec:Poisson}

Motivated by queue models for roads such as in~\citet{vandaele2000queueing}, we simulate road data by considering the roads as buffers in which traffic is injected following an i.i.d.\ Poisson distributed random variable with mean $\lambda$, i.e.,
\begin{equation}
 \textnormal{Pr}\!\left(X_{(1),i}=x\right)=e^{-\lambda}\frac{\lambda^x}{x!},
\end{equation}
for the incoming traffic at node $\bb{X}_{(1)}$. Each car proceeds independently of the others and the traffic propagates through the network. Finally, a random Poisson noise with mean $\mu_{(j)}$ is independently added at each sensor $\bb{X}_{(j)}$; the noise represents unobserved cars from side roads or instrumental error of the sensors. For each case, we generated $n=10^6$ samples.

Assume that the maximum time distance between sensors is one. In real scenarios, if the sensors are too close, it may happen that the same car appears in two consecutive sensors in one time slot.
By the definition of the directed information~\eqref{eq:DI_def}, such instantaneous events are accounted for in the information flowing in both directions.
To verify this, in the synthetic model we also investigate the case where the cars can only appear in two successive sensors with one step delay, i.e., removing any instantaneous effect.

\begin{figure}
\centering
\begin{subfigure}[t]{0.35\textwidth}
\centering
\includegraphics[width=.85\linewidth]{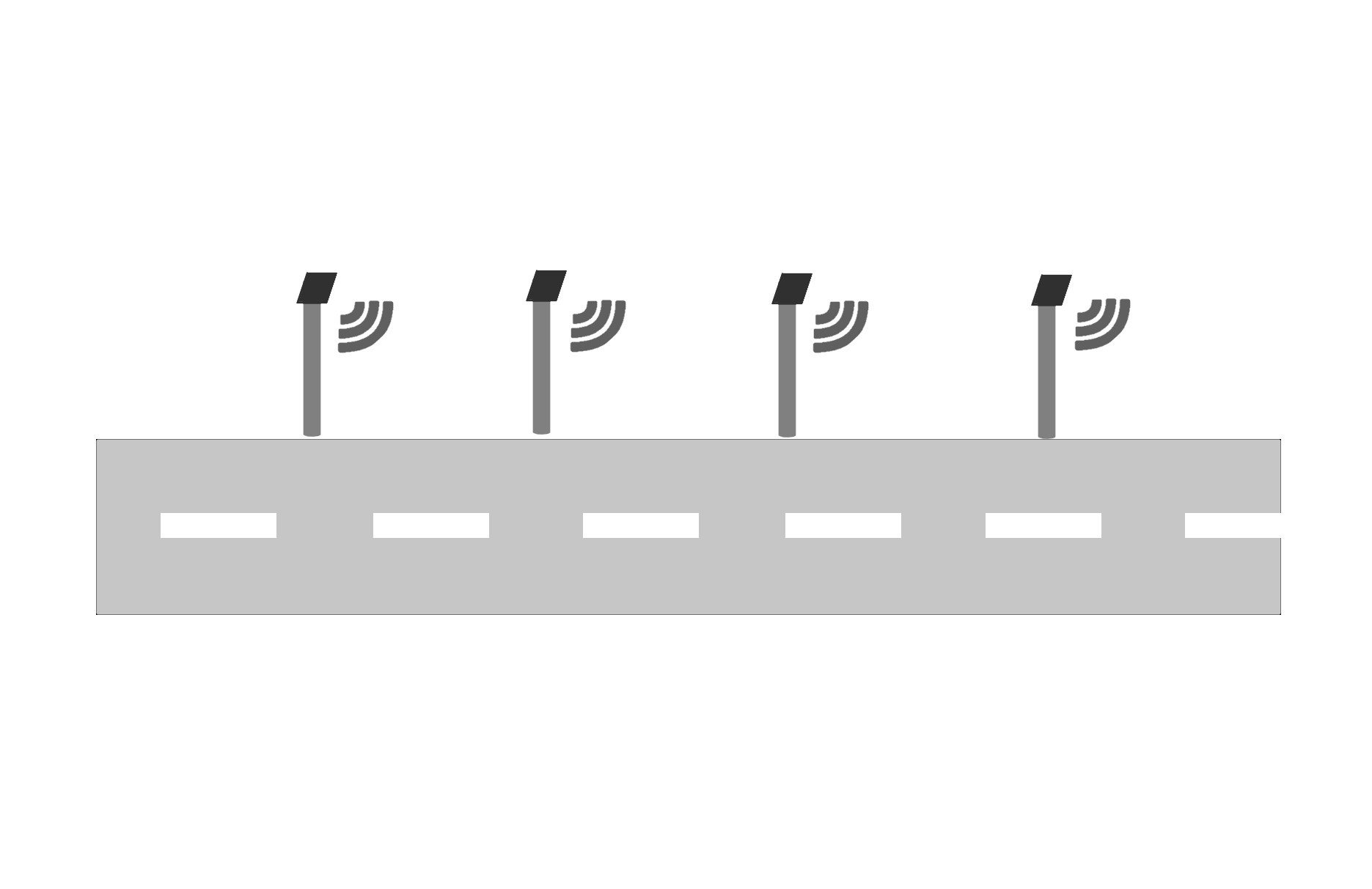}
\caption{Traffic on a single road.}
\label{fig:Sim_Row}
\end{subfigure}
~\quad
\begin{subfigure}[t]{0.35\textwidth}
\centering
\includegraphics[width=.85\linewidth]{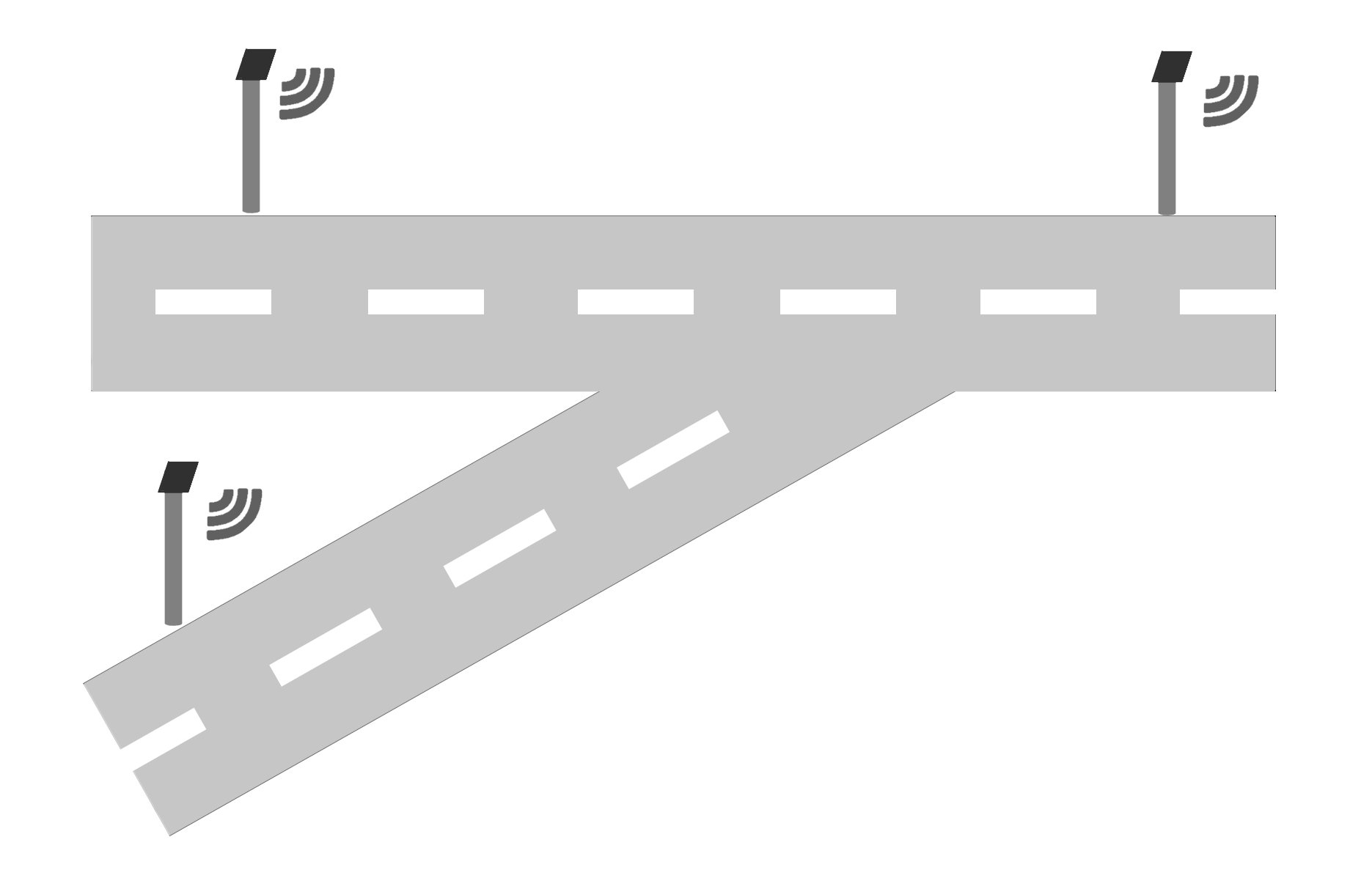}
\caption{Merging traffic scenario.}
\label{fig:Sim_Merge}
\end{subfigure}
\caption{Synthetic models.}
\end{figure}

\textbf{Sensors in sequence:}
In this setup, four sensors are located consecutively with equal distance.
To model both the rush hour and the low traffic period of a normal road, the mean of the input traffic alternates every 20 samples between $\lambda=5$ and $\lambda=1$ periodically. The noise parameters are $\mu_{(j)}=1,\, \forall j$. 
In the first scenario (S-I), each car appears at the next sensor with one frame delay, while in the second scenario (S-II), a car is allowed to appear simultaneously in two consecutive sensors.

\textbf{Merging traffic:}
In this scenario, two flows merge into one; we consider here three sensors as in Figure~\ref{fig:Sim_Merge}. The injected traffic ($\bb{X}_{(1)}$, $\bb{X}_{(2)}$) is assumed to be two independent Poisson random variables with means alternating periodically between $5$ and $1$, and the mean for the additive noise is set to be $\mu_{(j)}=1, \, \forall j$. 
Furthermore, by allowing only the cars from $\bb{X}_{(1)}$ to reach $\bb{X}_{(3)}$ instantaneously (less than the sampling time), we expect to detect flow of information in the backward direction $\bb{X}_{(3)}\to \bb{X}_{(1)}$ due to the definition~\eqref{eq:DI_def} (see also the discussion therein).
In a real traffic scenario, besides this instantaneous effect, a backward flow (with respect to the direction of traffic on the road) is expected to appear also when there is a congestion. However, in order to have a better control on the queue model and avoid complexity, we did not consider capacity for the link and model the congestion.

\begin{table}
\caption{Estimated $G_\nor$ for Poisson queue model using Algorithm~\ref{alg:DIG}.}
\label{tab:res_Simul}
\centering
\begin{tabular}{c | c | c}
S-I & S-II & S-III \\ \hline
\rule{0pt}{40pt}
$\begin{bmatrix}\cg{0} & \mathbf{1} & \cg{0} & \cg{0}\\ \cg{0} & \cg{0} & \mathbf{0.9}& \cg{0} \\ \cg{0} & \cg{0} & \cg{0}& \mathbf{1} \\ \cg{0} & \cg{0} & \cg{0}& \cg{0} \end{bmatrix}$
&$\begin{bmatrix}\cg{0} & \mathbf{1} & \cg{0.1} & \cg{0}\\\mathbf{0.6} & \cg{0} & \mathbf{0.4}& \cg{0} \\ \cg{0} & \cg{0.2} & \cg{0}& \mathbf{0.5} \\ \cg{0} & \cg{0} & \cg{0.3}& \cg{0} \end{bmatrix}$
& $\begin{bmatrix}\cg{0} & \cg{0.1} & \mathbf{0.7}\\ \cg{0.1} & \cg{0} & \mathbf{1}\\ \mathbf{0.5} & \cg{0.1} & \cg{0}\end{bmatrix}$
\end{tabular}
\end{table}

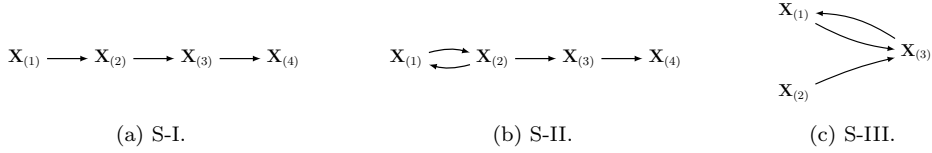
\begin{figure}[t]
\centering
\begin{subfigure}[t]{0.3\textwidth}
	\resizebox{\columnwidth}{!}{
	\begin{tikzpicture}[]
	\tikzstyle{nod} = [draw=none,fill=none,right]

	\node (T) [nod] at (0,1) {};
	\node (T2) [nod] at (0,-1) {};
	\node (S1) [nod] at (0,0) {$\bb{X}_{(1)}$};
	\node (S2) [nod] at (2,0) {$\bb{X}_{(2)}$};
	\node (S3) [nod] at (4,0) {$\bb{X}_{(3)}$};
	\node (S4) [nod] at (6,0) {$\bb{X}_{(4)}$};
	\draw [thick,-latex]  (S1) -- (S2);
	\draw [thick,-latex]  (S2) -- (S3);
	\draw [thick,-latex]  (S3) -- (S4);
	\end{tikzpicture}}
	\caption{S-I.}
\end{subfigure}
\qquad
\begin{subfigure}[t]{0.3\textwidth}
	\resizebox{\columnwidth}{!}{
	\begin{tikzpicture}[]
	\tikzstyle{nod} = [draw=none,fill=none,right]

	\node (T) [nod] at (0,2) {};
	\node (T2) [nod] at (0,0) {};
	\node (S1) [nod] at (0,1) {$\bb{X}_{(1)}$};
	\node (S2) [nod] at (2,1) {$\bb{X}_{(2)}$};
	\node (S3) [nod] at (4,1) {$\bb{X}_{(3)}$};
	\node (S4) [nod] at (6,1) {$\bb{X}_{(4)}$};
	\draw [thick,-latex]  (S1) to[bend left=15] (S2);
	\draw [thick,-latex]  (S2) to[bend left=15] (S1);
	\draw [thick,-latex]  (S2) -- (S3);
	\draw [thick,-latex]  (S3) -- (S4);
	\end{tikzpicture}}
	\caption{S-II.}
\end{subfigure}
\begin{subfigure}[t]{0.3\textwidth}
	\centering
	\resizebox{0.55\columnwidth}{!}{
	\begin{tikzpicture}[]
	\tikzstyle{nod} = [draw=none,fill=none,right]
	
	\node (S1) [nod] at (0,1) {$\bb{X}_{(1)}$};
	\node (S2) [nod] at (0,-1) {$\bb{X}_{(2)}$};
	\node (S3) [nod] at (3,0) {$\bb{X}_{(3)}$};
	\draw [thick,-latex]  (S1) to[bend right=10] (S3);
	\draw [thick,-latex]  (S2) to[bend left=5] (S3);
	\draw [thick,-latex]  (S3) to[bend right=15] (S1);
	\end{tikzpicture}}
	\caption{S-III.}
	\label{fig:DIG_Simul_Merge}
\end{subfigure}

\caption{Estimated DIG from synthetic data (Poisson model) with $d=1$ and threshold $\alpha=0.4$.}
\label{fig:DIG_Simul}
\end{figure}

\begin{figure}[t]
	\centering
	\includegraphics[width=\textwidth]{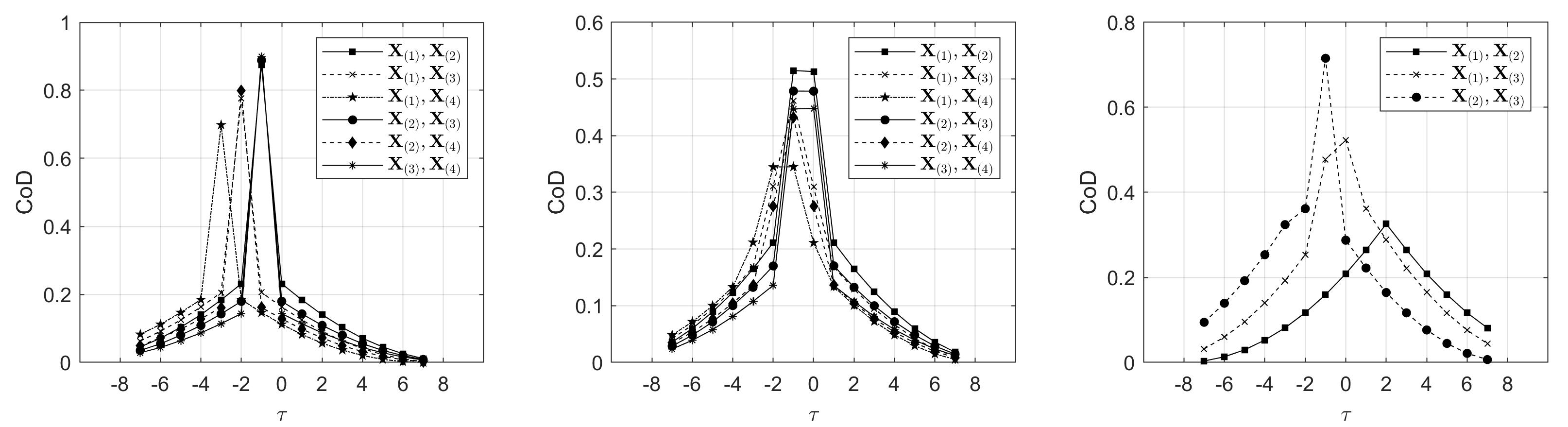}
	\caption{The computed CoD for each pair of nodes and different time shifts $\tau$, corresponding to scenarios S-I, S-II and S-III respectively from left.}
	\label{fig:COD}
\end{figure}

The graphs are estimated using Algorithm~\ref{alg:DIG} with the context tree estimator and two levels of quantization; the adjacency matrix $G$ and $G_\nor$ are computed.
Table~\ref{tab:res_Simul} states the normalized matrix $G_\nor$, with stronger links indicated with boldface (values above or equal to $\alpha=0.4$).
Finally the DIG is depicted for each scenario in Figure~\ref{fig:DIG_Simul}.
It can be observed how the instantaneous effect changes the flow of information in all links.
Note that in S-II a car can be fast enough to be observed at all sensors in one frame, while in S-III, only one link allows fast cars and that is the link where we observe a backward flow.

To complete our discussion, the values of CoD, defined as
$$CoD_{\bb{X}_{(i)},\bb{X}_{(j)}}(\tau)=\left[\frac{\bE{(X_{(i),t} -\mu_{\bb{X}_{(i)}})(X_{(j),t+\tau} -\mu_{\bb{X}_{(j)}})}}{\sigma_{\bb{X}_{(i)}} \sigma_{\bb{X}_{(j)}}}\right]^2,$$
are computed for each pair of nodes in Figure~\ref{fig:COD}. $\mu_{\bb{X}}$ and $\sigma_{\bb{X}}$ denote mean and standard deviation of $\bb{X}$ respectively. The location of peak indicates the time shift in which two traffic flows are alike, while the value of peak determines similarity in the corresponding time shift. However, It is non-trivial to obtain the causality and extent of the effect from one node to another by the values of CoD. Additionaly, CoD does not isolate the effect from other nodes in the transportaion network and exhibits only a pairwise relation (see the discussion corresponding to Figure~\ref{fig:effects}) 

\subsection{Synthetic data (CTM model)}
\label{Sec:CTM}

We consider the scenario of sensors on a single road (C-I) and merging traffic (C-II). 
The traffic dynamics are modeled according to a modification of the well-known Cell Transmission Model (\cite{daganzo1994cell}), 
where we divide the road into a number of cells and track the evolution of traffic density in each of them,
\begin{equation}
\label{eq:CTMPHI}
\rho_{i,l}(t+1) = \rho_{i,l}(t) + \frac{T}{L}\left( \Phi_{i,l}^+(t) - \Phi_{i,l}^-(t)\right), \quad i\in\{1: N_l\}, \ l\in\{1: K\}.
\end{equation}
Here $\rho_{i,l}(t)$ is the traffic density in cell $i$ of link $l$ at time $t$, $L$ the cell length, $T$ is the length of the time step, $N_l$ the number of cells in link $l$, $K$ the number of links in the considered road network, and $\Phi_{i,l}^+(t)$ and $\Phi_{i,l}^-(t)$ are the total flow during one time step into and out of cell $i$ of link $l$, respectively.
In scenario (C-I), we consider $K=1$ link with $N_1=100$ cells, and in scenario (C-II), we have $K=2$, $N_1=200$, $N_2=100$, and link $2$ merges into cell ${C_{2\rightarrow 1}=100}$ of link $1$.

The flows between cells are given by
\begin{align}
\Phi_{i,l}^-(t) = \Phi_{i+1,l}^+(t) &= \min\left(D_{i,l}(t), S_{i+1,l}(t)\right),\nonumber\\
D_{i,l}(t) &= \min\left(V_{i,l}(t)\rho_{i,l}(t), Q_{i,l}^{\max}\right),\nonumber\\
S_{i,l}(t) &= \min\left(W_{i,l}(t)\left(P_{i,l} - \rho_{i,l}(t)\right), Q_{i,l}^{\max}\right),
\end{align}
for $(i,l)$, $i\in\{1: N_l\}$, except for the two cells before and one cell after the merge, in which case we have
\begin{align}
\Phi_{C_{2\rightarrow 1}-1,1}^-(t) &= \min\left(D_{C_{2\rightarrow 1}-1,1}(t), S_{C_{2\rightarrow 1},1}(t)\frac{\rho_{C_{2\rightarrow 1}-1,1}(t)}{\rho_{C_{2\rightarrow 1}-1,1}(t)+\rho_{N_2,2}(t)}\right),\nonumber\\
\Phi_{N_2,2}^-(t) &= \min\left(D_{N_2,2}(t), S_{C_{2\rightarrow 1},1}(t)\frac{\rho_{N_2,2}(t)}{\rho_{C_{2\rightarrow 1}-1,1}(t)+\rho_{N_2,2}(t)}\right),\nonumber\\
\Phi_{C_{2\rightarrow 1},1}^+(t) &= \Phi_{C_{2\rightarrow 1}-1,1}^-(t) + \Phi_{N_2,2}^-(t),
\end{align}
effectively assuming proportional priority is assumed to the two merging roads.
The flow into the first cells of each link, $\Phi_{1,l}^+$ is defined externally.
In order to add stochasticity to the model, free flow speed $V_{i,l}(t)$ and congestion wave speed $W_{i,l}(t)$ of each cell are taken as uniformly distributed random variables: ${V_{i,l}(t) \sim \mathcal{U}(V_{\min}, V_{\max})}$ and ${W_{i,l}(t) \sim \mathcal{U}(W_{\min}, W_{\max})}$.

\begin{figure}[t]
\centering
\begin{subfigure}[t]{0.47\textwidth}
\centering
\includegraphics[width=0.9\textwidth]{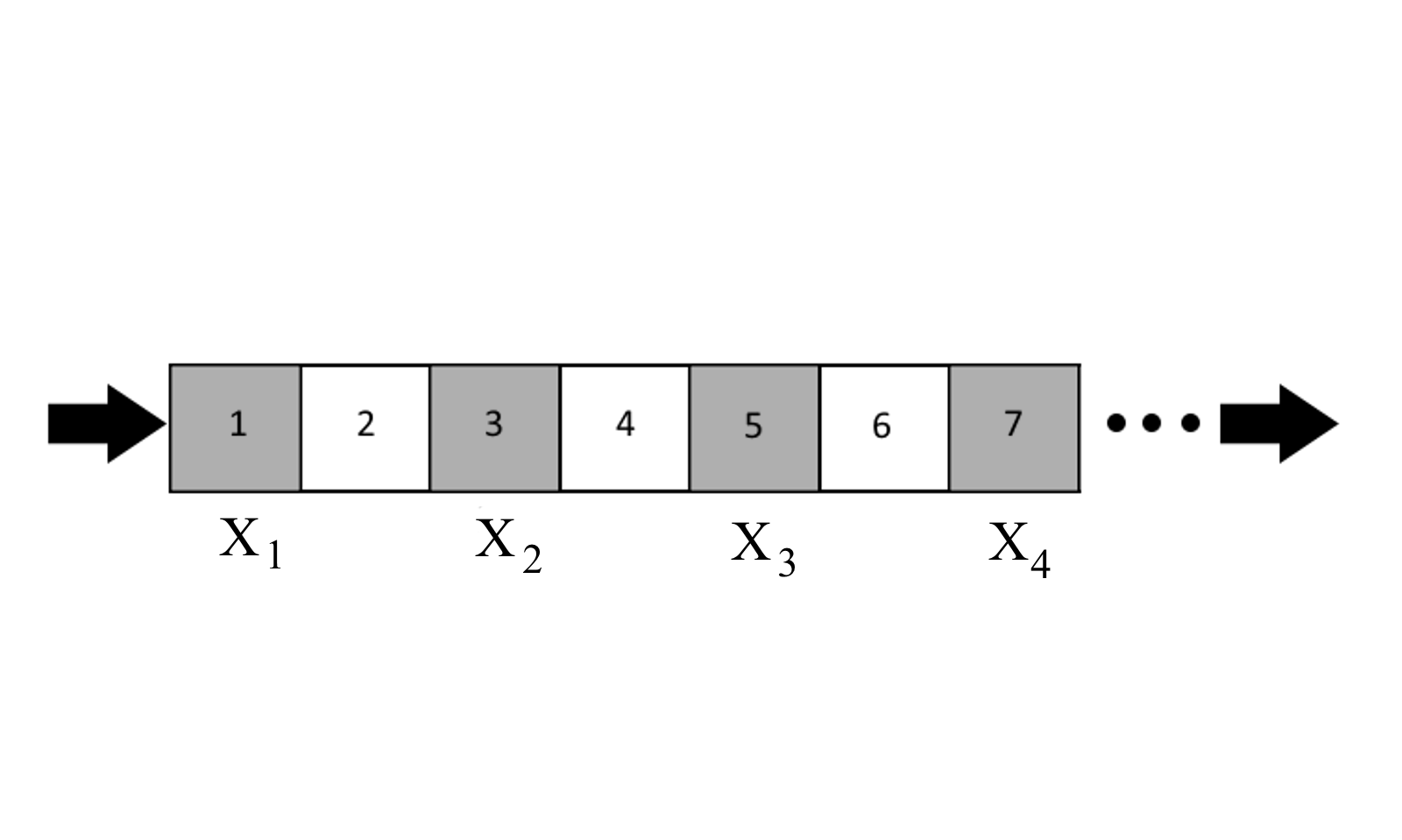}
\caption{Sensors on a row (C-I).}
\label{Fig:CTM_row}
\end{subfigure}\quad
\begin{subfigure}[t]{0.47\textwidth}
\centering
\includegraphics[width=0.7\textwidth]{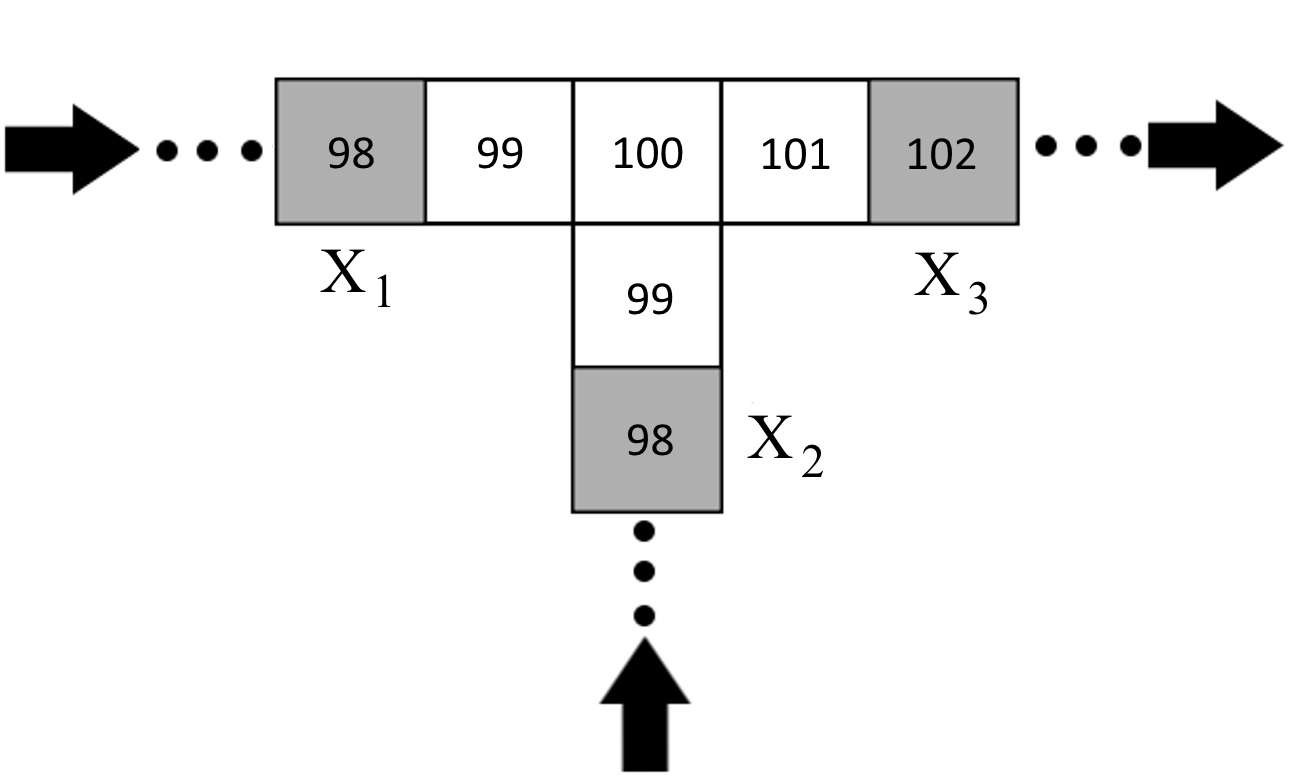}
\caption{Model for merging traffic (C-II).}
\label{Fig:CTM_merge}
\end{subfigure}
\caption{The CTM model for scenarios of a sequential sensors and merging traffic. Sensors are mounted at indicated cells. }
\end{figure}

\begin{table}[t]
\centering
\caption{Estimated matrix $G_\nor$ for scenarios C-I and C-II, with $d=6$ and $d=4$, respectively.}
\label{tab:res_CTM}
\begin{tabular}{c | c }
C-I & C-II \\ \hline
\rule{0pt}{50pt}
$\begin{bmatrix}\cg{0} & \mathbf{1} & \cg{0.6} & \cg{0.6}\\ \cg{0.6} & \cg{0} & \mathbf{0.8} & \cg{0.6} \\ \cg{0.5} & \cg{0.6} & \cg{0}& \mathbf{0.9} \\ \cg{0.4} & \cg{0.6} & \cg{0.6}& \cg{0} \end{bmatrix}$
&$\begin{bmatrix}\cg{0} &\cg{0.6}& \mathbf{1} & \\ \cg{0.5} & \cg{0} & \mathbf{1} \\ \cg{0.5} & \cg{0.6} & \cg{0}\end{bmatrix}$
\end{tabular}
\end{table}

Among all cells in scenario C-I we have chosen four at the beginning of the link to avoid the saturation due to the congestion which is propagating backward (Figure~\ref{Fig:CTM_row}). On the other hand, in C-II, we are interested in seeing the behavior at the merging point and three sensors are chosen accordingly (Figure~\ref{Fig:CTM_merge}).
We simulated $n=10^4$ samples for the two scenarios with the CT method and estimated the corresponding adjacency matrices using Algorithm~\ref{alg:DIG}. 
We obtained the DIGs with similar $G_\nor$ as in the Poisson model (Table~\ref{tab:res_CTM}).

\begin{figure}[t]
\centering
\includegraphics[width=\linewidth]{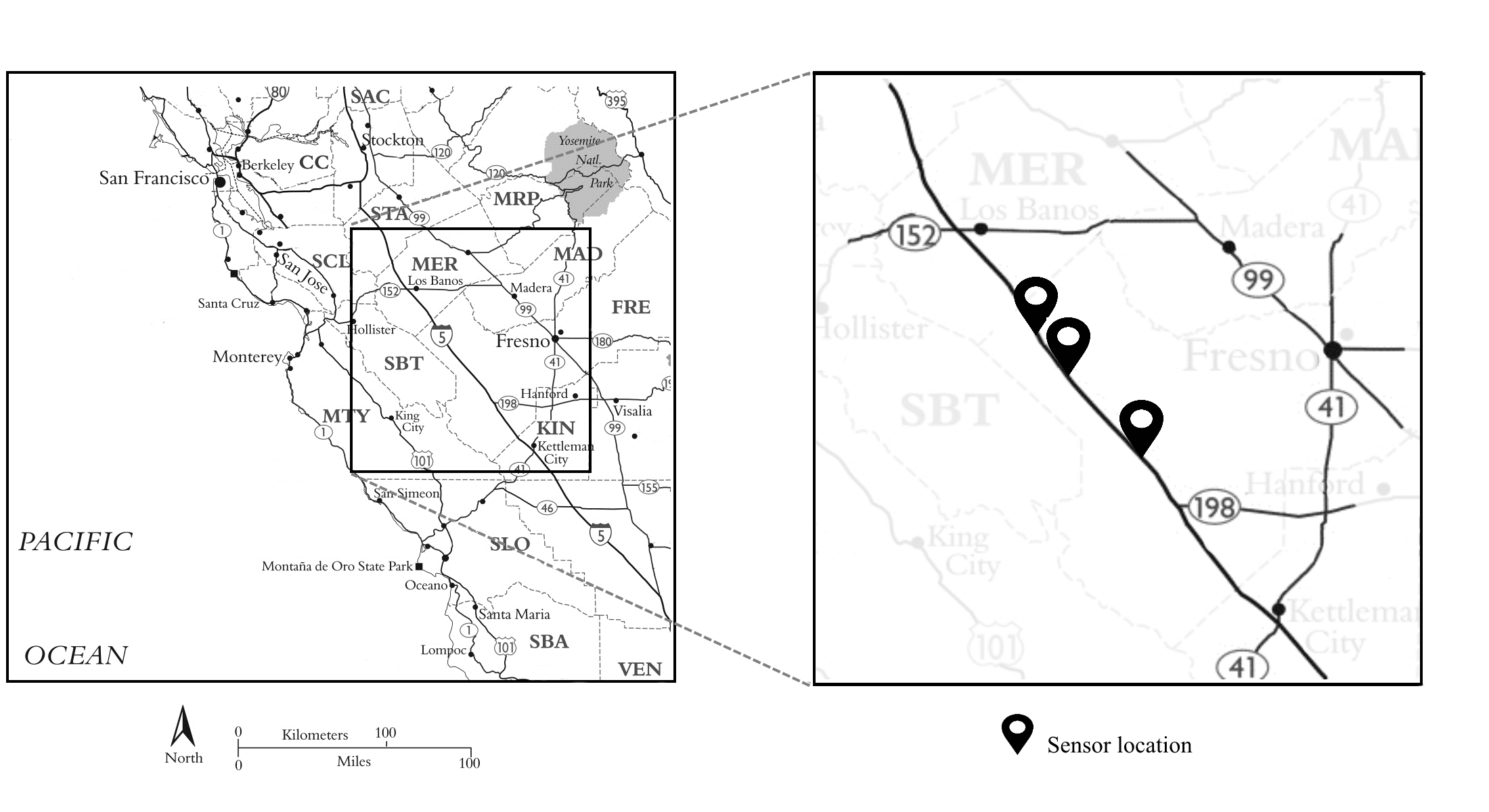}
\label{fig:Real_Row}
\caption{Sensors installed on interstate 5 south, Fresno, CA.}
\end{figure}

\begin{figure}[t]
\centering
\includegraphics[width=\linewidth]{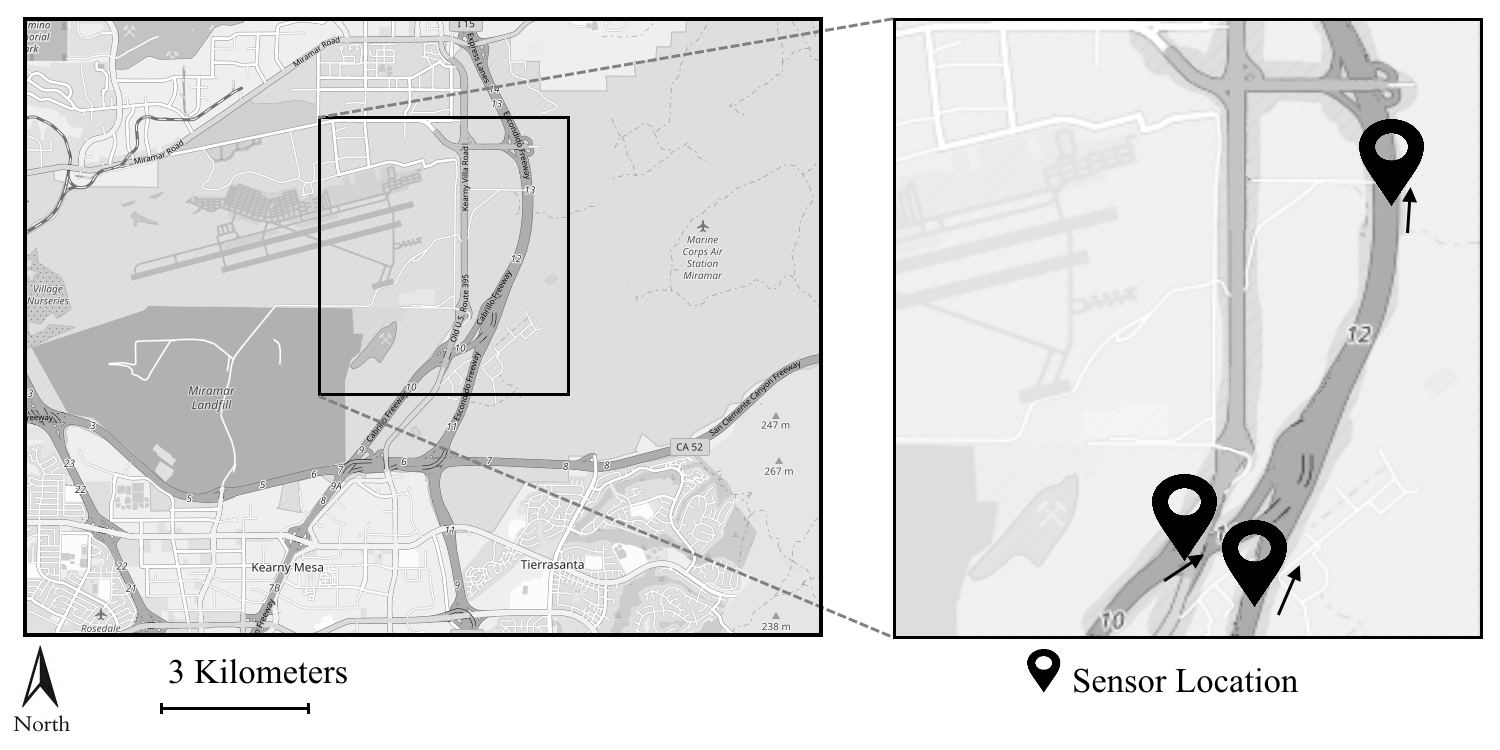}
\label{fig:Real_Merge}
\caption{Sensors installed on interstate 15 north and road 163 north, San Diego, CA.}
\end{figure}

\subsection{Real data}
\label{Sec:real}

To further test our estimator, traffic data for three different scenarios from the California Department of Transportation~(Caltrans)\footnote{Available at \url{http://pems.dot.ca.gov}.} is used. The data is aggregated traffic flow for every 5 minutes, extracted from installed sensors at the interstate 5 south in Fresno (R-I), the interstate 15 north in San Diego (R-II), and the downtown area of San Jose (R-III). 
We collected data from January 2017 to June 2019, every day, 24 hours. To avoid missing the cars that change lane, the aggregated data of all lanes in one side of the road is considered.

In the first setup, three sensors are chosen sequentially on the interstate 5 south, Fresno, CA such that, with an average speed, the distances between the sensors are approximately 5 and 25 minutes, respectively. As a result, the suggested depth to detect all causal links is $d=6$. However, estimated depth according to Algorithm~\ref{alg:DIG} is $d=5$.

In the second setup, we selected three sensors on a merging traffic scenario in San Diego, where the highway 160 north merges into the interstate 15 north. Despite the distance between sensors being approximately 5 minutes, the estimated depth obtained by cross-correlation is $d=6$.

\begin{table}[t]
\centering
\caption{Estimated matrix $G_\nor$ for scenarios R-I and R-II, with $d=5$ and $d=6$, respectively.}
\label{tab:Real}
\begin{tabular}{c | c}
R-I & R-II \\ \hline
\rule{0pt}{50pt}
$\begin{bmatrix}\cg{0} & \mathbf{1} & \cg{0.6} \\ \cg{0.6} & \cg{0} & \mathbf{0.7} \\ \cg{0.5} & \cg{0.5} & \cg{0}\end{bmatrix}$
&$\begin{bmatrix}\cg{0} & \cg{0.3} & \mathbf{1}\\ \cg{0.3} & \cg{0} & \mathbf{0.7} \\ \cg{0.5} & \cg{0.4} & \cg{0}\end{bmatrix}$
\end{tabular}
\end{table}

Table~\ref{tab:Real} contains the matrix $G_\nor$ estimated by the context tree method.
The initial hypothesis for causal structure for R-I is having two directed link $1\to 2$ and $2\to 3$, while for R-II we expect the main effects to be from sensors $1\to 3$ and $2\to 3$.
The estimated adjacency matrices confirm the initial hypothesis for both cases with a proper choice of the threshold.
However, $G_\nor$ reveals a non-negligible causal effect in other links that could be explained with the following ad hoc hypothesis. The existence of vehicles that appear in several sensors in the period of $d$ time steps (instantaneous observations in the controlled simulated scenarios S-II and S-III) results in a bidirectional dependency among sensors which appears in the values of the directed information. By changing the notion of causality to be strict, we may isolate such effects, forming a potential future research direction.

\begin{figure}[t]
\centering
\includegraphics[width=0.8\linewidth]{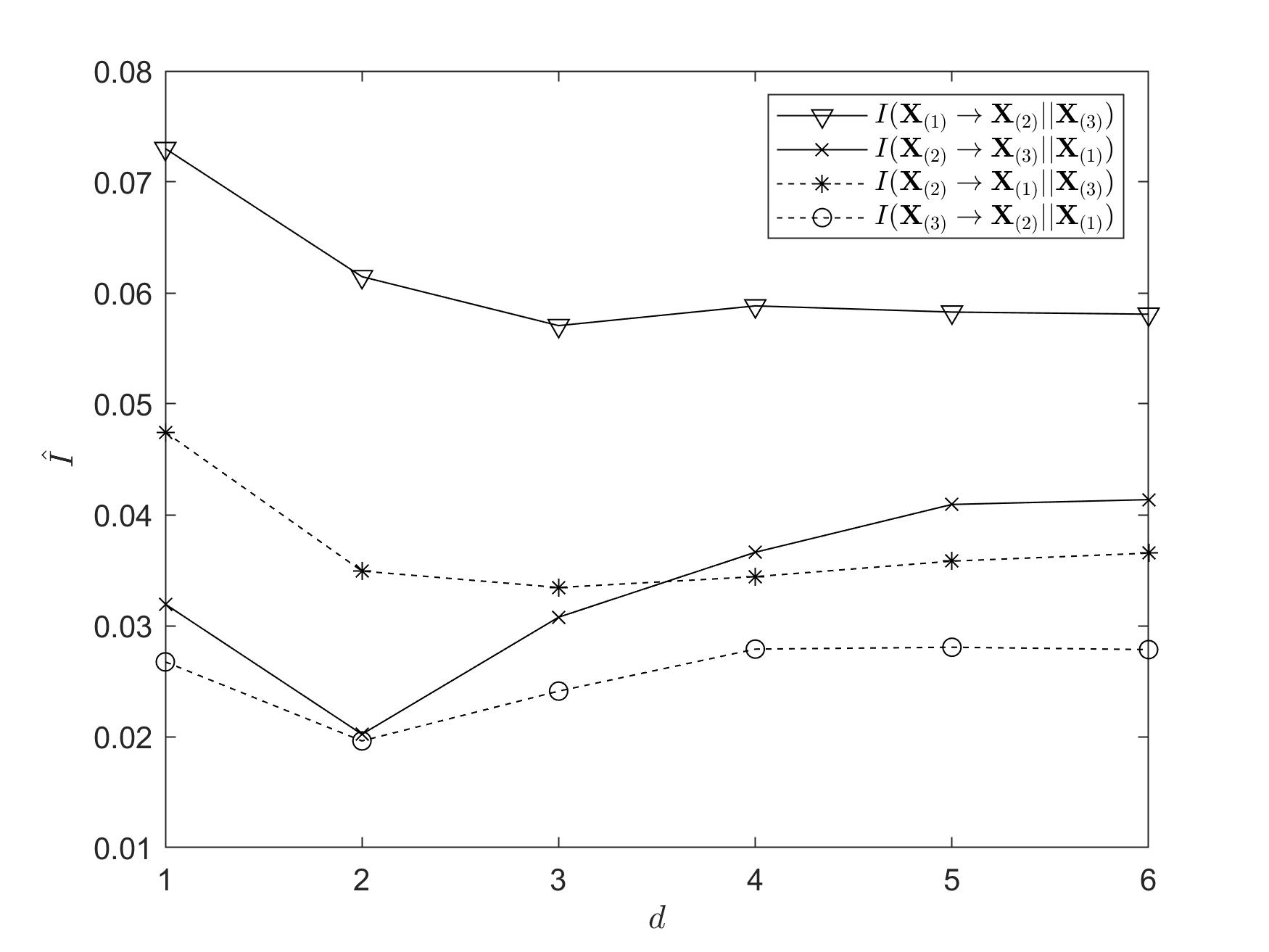}
\caption{Estimated values of the directed information for different choices of model depth $d$ in scenario R-I.}
\label{fig:depth}
\end{figure}

In Figure~\ref{fig:depth}, the values of $I(\co{\bb{X}_{(i)}\to \bb{X}_{(j)}}{\bb{X}_{\{1,2,3\}\setminus\{i,j\}}})$ for the scenario R-I are shown for different choices of $d$ and $i,j$.
We note that all the values seem to have converged when $d=6$ which validates the estimation we have used for the depth of the Markov model. 
Furthermore, the trend of values before convergence can be explained by the discussion in Section~\ref{Subsec:Est_depth} for an insufficient memory depth of the estimator.

\begin{figure}[t]
	\centering
	\includegraphics[width=\linewidth]{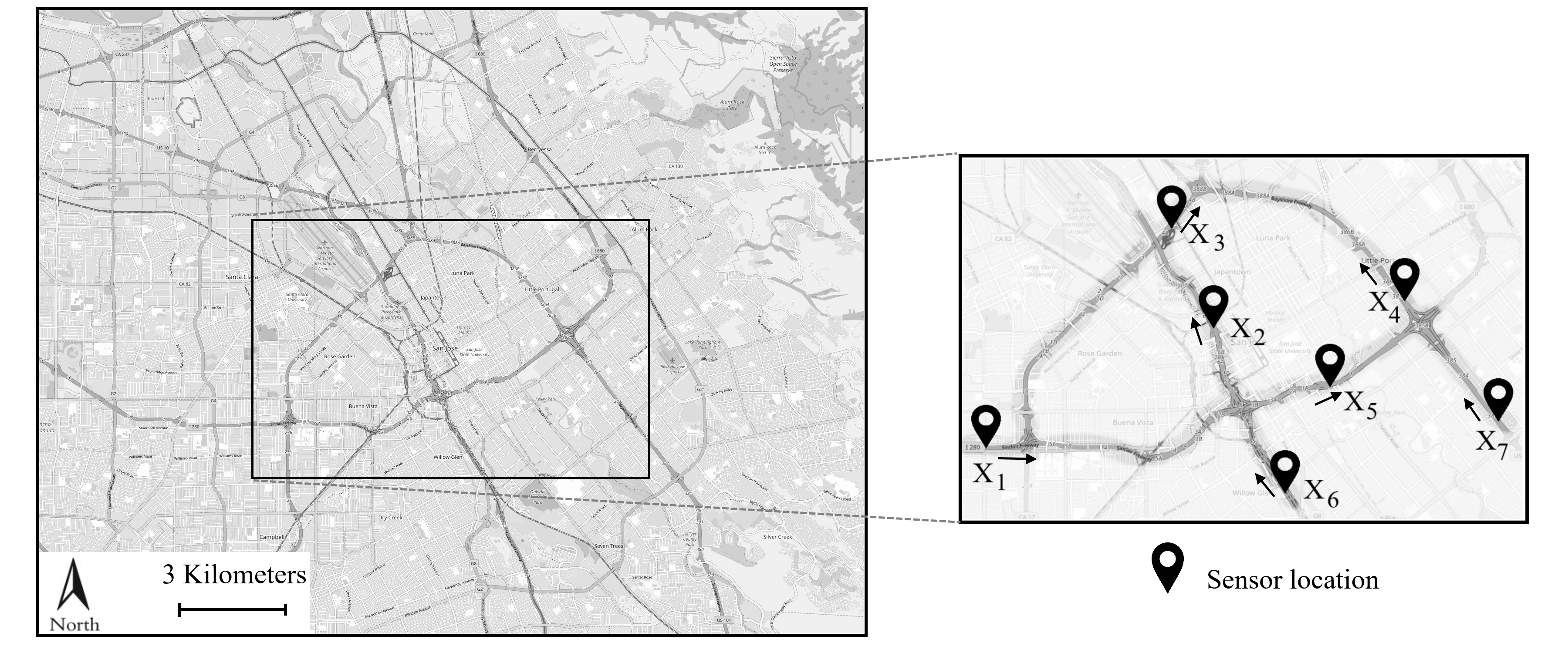}
	\caption{Sensors installed on highways in the downtown area of San Jose, CA. An arrow indicates physical direction of the lane(s) where a sensor is mounted. Note that there is no direct path from $\bb{X}_{(2)}$ to $\bb{X}_{(3)}$ since the major traffic from $\bb{X}_{(2)}$ passes above a bridge.}
    \label{fig:Real_Mix}
\end{figure}

\begin{figure}[t]
	\centering
	\begin{subfigure}[t]{0.3\textwidth}
		\resizebox{\columnwidth}{!}{
			\begin{tikzpicture}[]
			\tikzstyle{nod} = [draw=none,fill=none,right]
			\node (S1) [nod] at (0,0) {$\bb{X}_{(1)}$};
			\node (S3) [nod] at (2,1) {$\bb{X}_{(3)}$};
			\node (S5) [nod] at (2,-1) {$\bb{X}_{(5)}$};
			\node (S4) [nod] at (3.75,-1) {$\bb{X}_{(4)}$};
			\node (S2) [nod] at (3.75,1) {$\bb{X}_{(2)}$};
			\node (S6) [nod] at (5.5,1) {$\bb{X}_{(6)}$};
			\node (S7) [nod] at (5.5,-1) {$\bb{X}_{(7)}$};
			\draw [thick,-latex]  (S1) -- (S3);
			\draw [thick,-latex]  (S1) -- (S5);
			\draw [thick,-latex]  (S5) -- (S4);
			\draw [thick,-latex]  (S6) -- (S2);
			\draw [thick,-latex]  (S7) -- (S4);
			\draw [thick,-latex]  (S1) to[bend right=10] (S2);
			\draw [thick,-latex]  (S6) -- (S5);
			\end{tikzpicture}}
		\caption{}
		\label{fig:SanJose_physical}
	\end{subfigure}
	\hspace{.3cm}
	\begin{subfigure}[t]{0.3\textwidth}
		\centering
		\resizebox{\columnwidth}{!}{
			\begin{tikzpicture}[]
			\tikzstyle{nod} = [draw=none,fill=none,right]
			\node (T) [nod] at (6,0) {};
			\node (S1) [nod] at (0,0) {$\bb{X}_{(1)}$};
			\node (S3) [nod] at (2,1) {$\bb{X}_{(3)}$};
			\node (S5) [nod] at (2,-1) {$\bb{X}_{(5)}$};
			\node (S4) [nod] at (3.75,-1) {$\bb{X}_{(4)}$};
			\node (S2) [nod] at (3.75,1) {$\bb{X}_{(2)}$};
			\node (S6) [nod] at (5.5,1) {$\bb{X}_{(6)}$};
			\node (S7) [nod] at (5.5,-1) {$\bb{X}_{(7)}$};
			\draw [thick,-latex]  (S1) -- (S3);
			\draw [thick,-latex]  (S1) -- (S5);
			\draw [thick,-latex]  (S2) -- (S6);
			\draw [thick,-latex]  (S4) -- (S7);
			\end{tikzpicture}}
		\caption{}
		\label{fig:DIG_SanJose_7s}
	\end{subfigure}
	\hspace{.3cm}
	\begin{subfigure}[t]{0.3\textwidth}
	\resizebox{\columnwidth}{!}{
		\begin{tikzpicture}[]
		\tikzstyle{nod} = [draw=none,fill=none,right]
		\node (T) [nod] at (6,0) {};
		\node (S1) [nod] at (0,0) {$\bb{X}_{(1)}$};
		\node (S3) [nod] at (2,1) {$\bb{X}_{(3)}$};
		\node (S5) [nod] at (2,-1) {$\bb{X}_{(5)}$};
		\node (S4) [nod] at (3.75,-1) {$\bb{X}_{(4)}$};
		\node (S2) [nod] at (3.75,1) {$\bb{X}_{(2)}$};
		\node (S6) [nod] at (5.5,1) {$\bb{X}_{(6)}$};
		\draw [thick,-latex]  (S1) -- (S3);
		\draw [thick,latex-latex]  (S1) -- (S5);
		\draw [thick,dashed,-latex]  (S4) -- (S6);
		\draw [thick,latex-latex]  (S6) -- (S2);
		\end{tikzpicture}}
	\caption{}
	\label{fig:DIG_SanJose_6s}
	\end{subfigure}	
	\caption{(a) shows the major connections of the roads in R-III. The DIG is estimated with $d=3$ and threshold $\alpha=0.7$ in (b) and (c). Sensor $\bb{X}_{(7)}$ is excluded in (c) to experiment effect of latent node.}
	\label{fig:DIG_SanJose}
\end{figure}
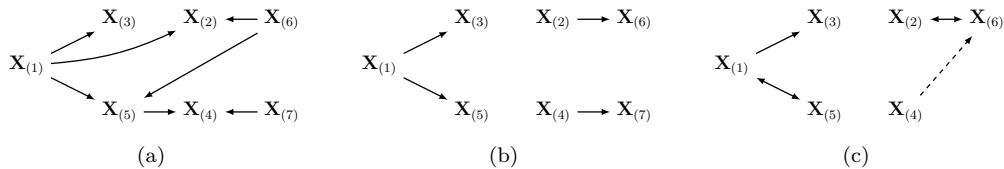

The last real-world scenario (R-III) is a combination of sensors mounted in San Jose, CA in a more complex road structure (Figure~\ref{fig:Real_Mix}). The most direct physical connections among sensors, depicted in Figure~\ref{fig:SanJose_physical} as a directed graph, are expected to be detected as causal links.
The DIG is estimated from the aggregated data using Algorithm~\ref{alg:DIG} with $\alpha=0.7$ and memory depth of $3$. The obtained graph for the whole network is depicted in Figure~\ref{fig:DIG_SanJose_7s} and shows that most of the expected links are indeed correctly detected.
The missing edges $\bb{X}_{(1)}\to\bb{X}_{(2)}$, $\bb{X}_{(5)}\to\bb{X}_{(4)}$ and $\bb{X}_{(6)}\to\bb{X}_{(5)}$ correspond to indirect highway connections so the links were not significant, while the backward links can be due to back propagation of downstream congestion.
Moreover, we have limited the number of sensors in order to provide a simple visual result; this, in turn, implies that not all the inbound and outbound traffic of the network is accounted for and it is possible to miss part of the flow, which may result in an inaccurate estimation of the causal links.  	
In fact, DIG heavily depends on the nodes included in the network. In other words, latent nodes which are excluded can affect the overall estimation due to the creation of fictitious causal links among other nodes. Such phenomena are explained as cascade and proxy effects in Section~\ref{sec:DIG}. To observe this, we designed the following experiment.

\subsubsection*{Latent node experiment}

The aim of this experiment is to show the effect of excluding the sensor $\bb{X}_{(7)}$ from the estimation. The obtained graph is depicted in Figure~\ref{fig:DIG_SanJose_6s} where we observe an unexpected link $\bb{X}_{(4)} \to \bb{X}_{(6)}$ (indicated with a dashed line). 
A possible explanation is that both $\bb{X}_{(6)}$ and $\bb{X}_{(7)}$ are mounted on the same direction on two parallel major highways, so a hidden variable such as rush hour flow or an unaccounted sensor upstream can make them dependent (i.e., share information) as in the cascade effect.
As a result, without considering $\bb{X}_{(7)}$, it is inferred that a link exists between $\bb{X}_{(4)}$ and $\bb{X}_{(6)}$, i.e., there is a flow of information.

The above experiment suggests that in order to have an accurate estimation for complex networks, it is necessary to include as many sensors as possible in the network. Although we have shown how directed information works as a measure of causality in synthetic scenarios and more isolated real world cases (R-I and R-II), with larger networks, the algorithm becomes computationally expensive and dealing with the consequent issues requires further investigation.

\section{Conclusion}
\label{sec:concl}
In this paper we have introduced a new method to study the causal effects of vehicular flow in traffic networks.
By estimating the causally conditioned directed information among the network nodes, the underlying graph is obtained; the intensity of the effects, given by the value of the directed information, determines the weights of each edge.
The results of this method on synthetic scenarios and real data collected from roads suggest that the directed information graph can be used to identify the underlying causal relations in a transportation network.

In this work the traffic flow was quantized with fixed thresholds; however, since the flow patterns differ depending on the time of the day, e.g., rush hours, a more advanced quantizer to capture all traffic fluctuations may be needed.
Moreover, the proposed algorithm, as it is presented here, becomes computationally expensive as the number of nodes increases and an improved version needs to be developed to efficiently deal with large sets of sensors.
Additionally, investigating the backward effect in congestion scenarios with the help of information flow requires a deeper study, which was out of the scope of this paper.

\section*{References}

\bibliography{ref}

\begin{thebibliography}{28}
\expandafter\ifx\csname natexlab\endcsname\relax\def\natexlab#1{#1}\fi
\providecommand{\url}[1]{\texttt{#1}}
\providecommand{\href}[2]{#2}
\providecommand{\path}[1]{#1}
\providecommand{\DOIprefix}{doi:}
\providecommand{\ArXivprefix}{arXiv:}
\providecommand{\URLprefix}{URL: }
\providecommand{\Pubmedprefix}{pmid:}
\providecommand{\doi}[1]{\href{http://dx.doi.org/#1}{\path{#1}}}
\providecommand{\Pubmed}[1]{\href{pmid:#1}{\path{#1}}}
\providecommand{\bibinfo}[2]{#2}
\ifx\xfnm\relax \def\xfnm[#1]{\unskip,\space#1}\fi
\bibitem[{Alexander et~al.(2015)Alexander, Jiang, Murga and
  Gonz\'{a}lez}]{ALEXANDER2015240}
\bibinfo{author}{Alexander, L.}, \bibinfo{author}{Jiang, S.},
  \bibinfo{author}{Murga, M.}, \bibinfo{author}{Gonz\'{a}lez, M.C.},
  \bibinfo{year}{2015}.
\newblock \bibinfo{title}{Origin–destination trips by purpose and time of day
  inferred from mobile phone data}.
\newblock \bibinfo{journal}{Transportation Research Part C: Emerging
  Technologies} \bibinfo{volume}{58}, \bibinfo{pages}{240 -- 250}.
\bibitem[{Amblard and Michel(2011)}]{amblard2011directed}
\bibinfo{author}{Amblard, P.O.}, \bibinfo{author}{Michel, O.J.},
  \bibinfo{year}{2011}.
\newblock \bibinfo{title}{On directed information theory and granger causality
  graphs}.
\newblock \bibinfo{journal}{Journal of computational neuroscience}
  \bibinfo{volume}{30}, \bibinfo{pages}{7--16}.
\bibitem[{Besselink et~al.(2016)Besselink, Turri, Van De~Hoef, Liang, Alam,
  M{\aa}rtensson and Johansson}]{besselink2016cyber}
\bibinfo{author}{Besselink, B.}, \bibinfo{author}{Turri, V.},
  \bibinfo{author}{Van De~Hoef, S.H.}, \bibinfo{author}{Liang, K.Y.},
  \bibinfo{author}{Alam, A.}, \bibinfo{author}{M{\aa}rtensson, J.},
  \bibinfo{author}{Johansson, K.H.}, \bibinfo{year}{2016}.
\newblock \bibinfo{title}{Cyber--physical control of road freight transport}.
\newblock \bibinfo{journal}{Proceedings of the IEEE} \bibinfo{volume}{104},
  \bibinfo{pages}{1128--1141}.
\bibitem[{Cai et~al.(2016)Cai, Wang, Lu, Chen, Ding and Sun}]{CAI201621}
\bibinfo{author}{Cai, P.}, \bibinfo{author}{Wang, Y.}, \bibinfo{author}{Lu,
  G.}, \bibinfo{author}{Chen, P.}, \bibinfo{author}{Ding, C.},
  \bibinfo{author}{Sun, J.}, \bibinfo{year}{2016}.
\newblock \bibinfo{title}{A spatiotemporal correlative k-nearest neighbor model
  for short-term traffic multistep forecasting}.
\newblock \bibinfo{journal}{Transportation Research Part C: Emerging
  Technologies} \bibinfo{volume}{62}, \bibinfo{pages}{21 -- 34}.
\bibitem[{Cai et~al.(2017)Cai, Neveu, Baxter, Byrne and
  Aazhang}]{cai2017inferring}
\bibinfo{author}{Cai, Z.}, \bibinfo{author}{Neveu, C.L.},
  \bibinfo{author}{Baxter, D.A.}, \bibinfo{author}{Byrne, J.H.},
  \bibinfo{author}{Aazhang, B.}, \bibinfo{year}{2017}.
\newblock \bibinfo{title}{Inferring neuronal network functional connectivity
  with directed information}.
\newblock \bibinfo{journal}{American Journal of Physiology-Heart and
  Circulatory Physiology} .
\bibitem[{Cheng et~al.(2012)Cheng, Haworth and Wang}]{cheng2012spatio}
\bibinfo{author}{Cheng, T.}, \bibinfo{author}{Haworth, J.},
  \bibinfo{author}{Wang, J.}, \bibinfo{year}{2012}.
\newblock \bibinfo{title}{Spatio-temporal autocorrelation of road network
  data}.
\newblock \bibinfo{journal}{Journal of Geographical Systems}
  \bibinfo{volume}{14}, \bibinfo{pages}{389--413}.
\bibitem[{Daganzo(1994)}]{daganzo1994cell}
\bibinfo{author}{Daganzo, C.F.}, \bibinfo{year}{1994}.
\newblock \bibinfo{title}{The cell transmission model: A dynamic representation
  of highway traffic consistent with the hydrodynamic theory}.
\newblock \bibinfo{journal}{Transportation Research Part B: Methodological}
  \bibinfo{volume}{28}, \bibinfo{pages}{269--287}.
\bibitem[{Diamantopoulos et~al.(2013)Diamantopoulos, Kehagias, K{\"o}nig and
  Tzovaras}]{diamantopoulos2013investigating}
\bibinfo{author}{Diamantopoulos, T.}, \bibinfo{author}{Kehagias, D.},
  \bibinfo{author}{K{\"o}nig, F.G.}, \bibinfo{author}{Tzovaras, D.},
  \bibinfo{year}{2013}.
\newblock \bibinfo{title}{Investigating the effect of global metrics in travel
  time forecasting}, in: \bibinfo{booktitle}{16th International IEEE Conference
  on Intelligent Transportation Systems (ITSC 2013)},
  \bibinfo{organization}{IEEE}. pp. \bibinfo{pages}{412--417}.
\bibitem[{Ermagun et~al.(2017)Ermagun, Chatterjee and
  Levinson}]{ermagun2017using}
\bibinfo{author}{Ermagun, A.}, \bibinfo{author}{Chatterjee, S.},
  \bibinfo{author}{Levinson, D.}, \bibinfo{year}{2017}.
\newblock \bibinfo{title}{Using temporal detrending to observe the spatial
  correlation of traffic}.
\newblock \bibinfo{journal}{PloS one} \bibinfo{volume}{12},
  \bibinfo{pages}{e0176853}.
\bibitem[{Granger(1969)}]{granger1969investigating}
\bibinfo{author}{Granger, C.W.}, \bibinfo{year}{1969}.
\newblock \bibinfo{title}{{Investigating Causal Relations by Econometric Models
  and Cross-Spectral Methods}}.
\newblock \bibinfo{journal}{Econometrica: Journal of the Econometric Society} ,
  \bibinfo{pages}{424--438}.
\bibitem[{Jiao et~al.(2013)Jiao, Permuter, Zhao, Kim and
  Weissman}]{jiao2013universal}
\bibinfo{author}{Jiao, J.}, \bibinfo{author}{Permuter, H.H.},
  \bibinfo{author}{Zhao, L.}, \bibinfo{author}{Kim, Y.H.},
  \bibinfo{author}{Weissman, T.}, \bibinfo{year}{2013}.
\newblock \bibinfo{title}{{Universal Estimation of Directed Information}}.
\newblock \bibinfo{journal}{IEEE Transactions on Information Theory}
  \bibinfo{volume}{59}, \bibinfo{pages}{6220--6242}.
\bibitem[{Kamarianakis and Prastacos(2005)}]{KAMARIANAKIS2005119}
\bibinfo{author}{Kamarianakis, Y.}, \bibinfo{author}{Prastacos, P.},
  \bibinfo{year}{2005}.
\newblock \bibinfo{title}{Space–time modeling of traffic flow}.
\newblock \bibinfo{journal}{Computers \& Geosciences} \bibinfo{volume}{31},
  \bibinfo{pages}{119 -- 133}.
\newblock \bibinfo{note}{Geospatial Research in Europe: AGILE 2003}.
\bibitem[{Keimer et~al.(2018)Keimer, Laurent-Brouty, Farokhi, Signargout,
  Cvetkovic, Bayen and Johansson}]{keimer2018information}
\bibinfo{author}{Keimer, A.}, \bibinfo{author}{Laurent-Brouty, N.},
  \bibinfo{author}{Farokhi, F.}, \bibinfo{author}{Signargout, H.},
  \bibinfo{author}{Cvetkovic, V.}, \bibinfo{author}{Bayen, A.M.},
  \bibinfo{author}{Johansson, K.H.}, \bibinfo{year}{2018}.
\newblock \bibinfo{title}{Information patterns in the modeling and design of
  mobility management services}.
\newblock \bibinfo{journal}{Proceedings of the IEEE} \bibinfo{volume}{106},
  \bibinfo{pages}{554--576}.
\bibitem[{Kontoyiannis and Skoularidou(2016)}]{kontoyiannis2016estimating}
\bibinfo{author}{Kontoyiannis, I.}, \bibinfo{author}{Skoularidou, M.},
  \bibinfo{year}{2016}.
\newblock \bibinfo{title}{{Estimating the Directed Information and Testing for
  Causality}}.
\newblock \bibinfo{journal}{IEEE Transactions on Information Theory}
  \bibinfo{volume}{62}, \bibinfo{pages}{6053--6067}.
\bibitem[{Ma et~al.(2017)Ma, Koutsopoulos, Ferreira and
  Mesbah}]{ma2017estimation}
\bibinfo{author}{Ma, Z.}, \bibinfo{author}{Koutsopoulos, H.N.},
  \bibinfo{author}{Ferreira, L.}, \bibinfo{author}{Mesbah, M.},
  \bibinfo{year}{2017}.
\newblock \bibinfo{title}{Estimation of trip travel time distribution using a
  generalized markov chain approach}.
\newblock \bibinfo{journal}{Transportation Research Part C: Emerging
  Technologies} \bibinfo{volume}{74}, \bibinfo{pages}{1--21}.
\bibitem[{Min and Wynter(2011)}]{MIN2011606}
\bibinfo{author}{Min, W.}, \bibinfo{author}{Wynter, L.}, \bibinfo{year}{2011}.
\newblock \bibinfo{title}{Real-time road traffic prediction with
  spatio-temporal correlations}.
\newblock \bibinfo{journal}{Transportation Research Part C: Emerging
  Technologies} \bibinfo{volume}{19}, \bibinfo{pages}{606 -- 616}.
\bibitem[{Molavipour et~al.(2017)Molavipour, Bassi and
  Skoglund}]{Mol2017TestforDIG}
\bibinfo{author}{Molavipour, S.}, \bibinfo{author}{Bassi, G.},
  \bibinfo{author}{Skoglund, M.}, \bibinfo{year}{2017}.
\newblock \bibinfo{title}{Testing for directed information graphs}, in:
  \bibinfo{booktitle}{2017 55th Annual Allerton Conference on Communication,
  Control, and Computing (Allerton)}, pp. \bibinfo{pages}{212--219}.
\bibitem[{Pierce and Haugh(1977)}]{PIERCE1977265}
\bibinfo{author}{Pierce, D.A.}, \bibinfo{author}{Haugh, L.D.},
  \bibinfo{year}{1977}.
\newblock \bibinfo{title}{Causality in temporal systems: Characterization and a
  survey}.
\newblock \bibinfo{journal}{Journal of Econometrics} \bibinfo{volume}{5},
  \bibinfo{pages}{265 -- 293}.
\bibitem[{Quinn et~al.(2011)Quinn, Coleman, Kiyavash and
  Hatsopoulos}]{quinn2011estimating}
\bibinfo{author}{Quinn, C.J.}, \bibinfo{author}{Coleman, T.P.},
  \bibinfo{author}{Kiyavash, N.}, \bibinfo{author}{Hatsopoulos, N.G.},
  \bibinfo{year}{2011}.
\newblock \bibinfo{title}{{Estimating the Directed Information to Infer Causal
  Relationships in Ensemble Neural Spike Train Recordings}}.
\newblock \bibinfo{journal}{Journal of computational neuroscience}
  \bibinfo{volume}{30}, \bibinfo{pages}{17--44}.
\bibitem[{Quinn et~al.(2015)Quinn, Kiyavash and Coleman}]{quinn2015directed}
\bibinfo{author}{Quinn, C.J.}, \bibinfo{author}{Kiyavash, N.},
  \bibinfo{author}{Coleman, T.P.}, \bibinfo{year}{2015}.
\newblock \bibinfo{title}{{Directed Information Graphs}}.
\newblock \bibinfo{journal}{IEEE Transactions on Information Theory}
  \bibinfo{volume}{61}, \bibinfo{pages}{6887--6909}.
\bibitem[{{Rissanen} and {Wax}(1987)}]{Rissanen1987}
\bibinfo{author}{{Rissanen}, J.}, \bibinfo{author}{{Wax}, M.},
  \bibinfo{year}{1987}.
\newblock \bibinfo{title}{Measures of mutual and causal dependence between two
  time series (corresp.)}.
\newblock \bibinfo{journal}{IEEE Transactions on Information Theory}
  \bibinfo{volume}{33}, \bibinfo{pages}{598--601}.
\bibitem[{Salamanis et~al.(2016)Salamanis, Kehagias, Filelis-Papadopoulos,
  Tzovaras and Gravvanis}]{salamanis2016managing}
\bibinfo{author}{Salamanis, A.}, \bibinfo{author}{Kehagias, D.D.},
  \bibinfo{author}{Filelis-Papadopoulos, C.K.}, \bibinfo{author}{Tzovaras, D.},
  \bibinfo{author}{Gravvanis, G.A.}, \bibinfo{year}{2016}.
\newblock \bibinfo{title}{Managing spatial graph dependencies in large volumes
  of traffic data for travel-time prediction}.
\newblock \bibinfo{journal}{IEEE Transactions on Intelligent Transportation
  Systems} \bibinfo{volume}{17}, \bibinfo{pages}{1678--1687}.
\bibitem[{Treiber and Kesting(2012)}]{treiber2012validation}
\bibinfo{author}{Treiber, M.}, \bibinfo{author}{Kesting, A.},
  \bibinfo{year}{2012}.
\newblock \bibinfo{title}{Validation of traffic flow models with respect to the
  spatiotemporal evolution of congested traffic patterns}.
\newblock \bibinfo{journal}{Transportation research part C: emerging
  technologies} \bibinfo{volume}{21}, \bibinfo{pages}{31--41}.
\bibitem[{Tympakianaki et~al.(2015)Tympakianaki, Koutsopoulos and
  Jenelius}]{TYMPAKIANAKI2015231}
\bibinfo{author}{Tympakianaki, A.}, \bibinfo{author}{Koutsopoulos, H.N.},
  \bibinfo{author}{Jenelius, E.}, \bibinfo{year}{2015}.
\newblock \bibinfo{title}{c-spsa: Cluster-wise simultaneous perturbation
  stochastic approximation algorithm and its application to dynamic
  origin–destination matrix estimation}.
\newblock \bibinfo{journal}{Transportation Research Part C: Emerging
  Technologies} \bibinfo{volume}{55}, \bibinfo{pages}{231 -- 245}.
\bibitem[{Vandaele et~al.(2000)Vandaele, Van~Woensel and
  Verbruggen}]{vandaele2000queueing}
\bibinfo{author}{Vandaele, N.}, \bibinfo{author}{Van~Woensel, T.},
  \bibinfo{author}{Verbruggen, A.}, \bibinfo{year}{2000}.
\newblock \bibinfo{title}{A queueing based traffic flow model}.
\newblock \bibinfo{journal}{Transportation Research Part D: Transport and
  Environment} \bibinfo{volume}{5}, \bibinfo{pages}{121--135}.
\bibitem[{Vlahogianni et~al.(2014)Vlahogianni, Karlaftis and
  Golias}]{VLAHOGIANNI20143}
\bibinfo{author}{Vlahogianni, E.I.}, \bibinfo{author}{Karlaftis, M.G.},
  \bibinfo{author}{Golias, J.C.}, \bibinfo{year}{2014}.
\newblock \bibinfo{title}{Short-term traffic forecasting: Where we are and
  where we’re going}.
\newblock \bibinfo{journal}{Transportation Research Part C: Emerging
  Technologies} \bibinfo{volume}{43}, \bibinfo{pages}{3 -- 19}.
\newblock \bibinfo{note}{Special Issue on Short-term Traffic Flow Forecasting}.
\bibitem[{Willems(1998)}]{willems1998context}
\bibinfo{author}{Willems, F.M.}, \bibinfo{year}{1998}.
\newblock \bibinfo{title}{The context-tree weighting method: Extensions}.
\newblock \bibinfo{journal}{IEEE Transactions on Information Theory}
  \bibinfo{volume}{44}, \bibinfo{pages}{792--798}.
\bibitem[{Willems et~al.(1995)Willems, Shtarkov and
  Tjalkens}]{willems1995context}
\bibinfo{author}{Willems, F.M.}, \bibinfo{author}{Shtarkov, Y.M.},
  \bibinfo{author}{Tjalkens, T.J.}, \bibinfo{year}{1995}.
\newblock \bibinfo{title}{The context-tree weighting method: basic properties}.
\newblock \bibinfo{journal}{IEEE Transactions on Information Theory}
  \bibinfo{volume}{41}, \bibinfo{pages}{653--664}.

\end{thebibliography}

%
%
%

\end{document}